\documentclass[useAMS,usenatbib]{mn2e}

\usepackage{graphicx} 
\usepackage[applemac]{inputenc}
\usepackage{epsfig}
\usepackage{comment}
\usepackage{natbib}
 \usepackage{lscape}
 \usepackage{colortbl}
 \usepackage{verbatim} 
\usepackage{url}
 \bibpunct{(}{)}{;}{a}{}{,}

\title{The evolution of the mass--size relation for
early type galaxies from $z\sim1$ to the present: dependence on environment, mass--range and detailed morphology}

\author[M. Huertas-Company  et al.]{M. Huertas-Company$^{1,2}$\thanks{E-mail:
marc.huertas@obspm.fr }, S. Mei$^{1,2}$, {F. Shankar$^{1}$}, L. Delaye$^{1,9}$, A. Raichoor$^{3}$,
\newauthor
G. Covone$^{4,5}$, {A. Finoguenov}$^{6}$, J.P. Kneib$^{7}$ , O. Le F\`evre$^{7}$, M. Povi\`c$^{8}$\ \\
$^{1}$GEPI, Paris Observatory, 77 av. Denfert Rochereau, 75014 Paris, France\\
$^{2}$University Denis Diderot, 4 Rue Thomas Mann, 75205 Paris, France\\
$^{3}$ INAF – Osservatorio Astronomico di Brera, via Brera 28, 20121 Milan, Italy \\
$^{4}$Dipartimento di Scienze Fisiche, Universit\`a di Napoli ''Federico II'', Via Cinthia, I-80126 Napoli, Italy \\
$^{5}$INFN Sez. di Napoli, Compl. Univ. Monte S. Angelo, Via Cinthia, I-80126 Napoli, Italy\\
$^{6}$Department of Physics, University of Helsinki, Gustaf Hällströmin katu2a, FI-00014 Helsinki, Finland\\
$^{7}$ Laboratoire d'Astrophysique de Marseille, CNRS-Universit\'e\\
$^{8}$ Instituto de astrof\'isica de Andaluc\'ia(IAA-CSIC), C/ Glorieta de la Astronom\'ia s/n, 18008 Granada, Spain\\
$^{9}$ CEA, IRFU, SAp, F-91191 Gif-sur-Yvette, France}

\begin{document}

\date{Accepted 1988 December 15. Received 1988 December 14; in original form 1988 October 11}

\pagerange{\pageref{firstpage}--\pageref{lastpage}} \pubyear{2002}

\maketitle

\label{firstpage}

\begin{abstract}

We study the dependence of the galaxy size evolution on morphology, stellar mass and large scale environment for a sample of 298 group and 384 field quiescent early-type galaxies from the COSMOS survey, selected  from $z\sim1$ to the present, and with masses $log(M/M_\odot)>10.5$. 

From a detailed morphological analysis we infer that $\sim80\%$ of passive galaxies with mass $log(M/M_\odot)>10.5$  have an early-type morphology and that this fraction does not evolve over the last 6 Gyr. However the relative abundance of lenticular and elliptical galaxies depends on stellar mass. Elliptical galaxies dominate only at the very high mass end --~$log(M/M_\odot)>11$~-- while S0 galaxies dominate at lower stellar masses  --~$10.5<log(M/M_\odot)<11$. 

The galaxy size growth depends on galaxy mass range and  early--type galaxy morphology, e.g., elliptical galaxies evolve differently than lenticular galaxies.  At the low mass end  --~$10.5<Log(M/M_\odot)<11$, ellipticals do not show strong size growth from $z\sim1$ to the present (10\% to 30\% depending on the morphological classification). On the other end, massive ellipticals --~$log(M/M_\odot)>11.2$~-- approximately doubled their size. 
Interestingly, lenticular galaxies display different behavior: they appear more compact on average and they do show a size growth of $\sim60\%$ since $z=1$ independent of stellar mass  range. 

We compare our results with state-of-the art semi-analytic models.  While major and minor mergers can account for most of the galaxy size growth, we find that with present data and the theoretical uncertainties in the modeling we cannot state clear evidence favoring either merger or mass loss via quasar and/or stellar winds as the primary mechanism driving the evolution.

The galaxy mass-size relation and size growth do not depend on environment
in the halo mass range explored in this work (field to group mass $log(M_h/M_\odot) < 14)$,
i.e., group and field galaxies follow the same trends. At low redshift, where we examine both SDSS and COSMOS groups, this result is at variance with
predictions from some current hierarchical models that show a clear dependence of size
growth on halo mass for massive ellipticals ( $log(M_*/M_\odot) > 11.2$). In future work we will analyze in detail if this result is specific of the observations and model used in this work.

BCG and satellite galaxies lie on the same mass-size relation, at variance with
predictions from hierarchical models, which predict that BCGs should have larger
sizes than satellites because they experience more mergers in groups over the halo
mass range probed.



\end{abstract}

\begin{keywords}
galaxies: evolution, galaxies: elliptical and lenticular, cD, galaxies: groups
\end{keywords}

\section{Introduction}

The fact that massive quiescent galaxies experienced a strong size evolution in the last 10 Gyrs is now a commonly accepted picture since first works on this topic were published \citep{2005ApJ...626..680D, 2006ApJ...650...18T}. 
Many independent groups using different datasets and selections have come up with similar conclusions, i.e. massive galaxies roughly doubled their size from $z\sim1$ and probably increased it by $3\sim5$ from $z\sim2$ (e.g. \citealp{2008ApJ...688...48V,2008ApJ...677L...5V,2008ApJ...687L..61B,2011ApJ...739L..44D,2012MNRAS.tmpL.424C}) even though there might be a population with larger sizes already in place at high redshift (e.g \citealp{2010MNRAS.401..933M,2011MNRAS.412.2707S})
As several works pointed out since the first publications came out, the result could be biased by a wrong estimate of stellar masses (which is usually done through SED fitting, e.g. \citealp{2011ApJ...732...12R}) and/or an under-estimate of galaxy sizes at very high redshifts since surface brightness dimming could cause the loss of the very outer parts of galaxies (e.g. \citealp{2009ApJ...697.1290B,2009MNRAS.398..898H}). Independent measurements based on dynamical masses have however confirmed the compactness of several objects (e.g. \citealp{2011ApJ...738L..22M, 2011ApJ...736L...9V, 2012ApJ...746..162N}) and it now seems clear that there is indeed a population of compact objects at high redshift. 
Some works also pointed out that we cannot exclude that the current galaxy selections might be biased and we are missing those compact objects in the nearby universe, or selecting the most compact objects at high redshift \citep{2010ApJ...712..226V}. This would be at variance with the work of \cite{2009ApJ...692L.118T} that showed that there is not a strong evidence for a significant fraction of compact objects in the local Universe.

From the theoretical point of view, two main mechanisms have been proposed to increase galaxy size. \cite{2008ApJ...689L.101F} proposed AGN and supernovae feedback as the main responsible of galaxy expansion while \cite{2006ApJS..163....1H} and \cite{2009ApJ...699L.178N} argued that minor dry mergers are the most efficient way to grow sizes since they affect the outer parts of the galaxy without significantly modifying the stellar mass nor the star formation (see also \citealp{2011arXiv1105.6043S}). 

Several recent observational works have tried to disentangle the two scenarios. As a result, minor dry merging seems to emerge as the most plausible explanation for the size evolution (e.g. \citealp{2011MNRAS.415.3903T}) at least from $z\sim1$, leading to an inside-out growth of galaxies (e.g. \citealp{2011MNRAS.411.1435T, 2012arXiv1208.0341P}) even though there is still some debate. \cite{2012ApJ...746..162N} showed by carefully counting small companions in very deep images from the CANDELS survey  \citep{2011arXiv1105.3753G, 2011ApJS..197...36K} that minor mergers are roughly enough to account for the size evolution from $z\sim1$ (provided that a short merger timescale is assumed) and a combination of minor merging and star formation quenching can explain the galaxy growth from $z\sim2$. Using morphological merging indicators, \cite{2011arXiv1111.5662B} reached a similar conclusion and even argue that the problem is close to be solved (see also \cite{2012arXiv1205.4058M}  for similar considerations). On the other hand, \cite{2011A&A...530A..20L,2012arXiv1202.4674L} point out that for galaxies with masses $log(M/M_\odot)>11$ minor mergers are not the only process responsible for size growth since $z \sim 1$. These authors propose a scenario for which minor and major mergers contribute to $\sim 55\%$, while the remaining $\sim 45\%$ to $\sim 25\%$ is due to other processes and specially to younger galaxies (hence larger) arriving at later epochs (progenitor bias). 

When measuring the galaxy mass--size relation and mass growth, different works use different sample selections though and this might lead to different conclusions. It is very rare in fact to find two works dealing with size evolution which apply the same criteria to select their galaxy sample and the same methodology to analyze the data. Some are based on star-formation only \citep{2012ApJ...750...93P} others on morphology \citep{2012ApJ...745..130R,2011arXiv1109.5698C} or on a combination of both (e.g. \citealp{2008ApJ...688...48V,2011ApJ...739L..44D}). Finally, many works combine a stellar mass selection with a quiescence criterion (e.g. \citealp{2008ApJ...677L...5V,2012ApJ...746..162N}) and generally the mass cuts are not always the same. See \cite{2011ApJ...739L..44D} and \cite{2012MNRAS.tmpL.424C} for two different compilations of recent results. These different selections are based on the general idea that almost all massive galaxies are passive and have an early-type morphology, which is not always true as shown in recent works (e.g. \citealp{2011ApJ...743L..15V, 2012ApJ...751...45T})

Another recent point of discussion in the literature is the role of environment. Recent observational studies show controversial results. 
\cite{2012ApJ...745..130R} studied the mass size relation for morphologically selected early-type galaxies at $z\sim1.2$ in three different environments (field, cluster, groups) and find that, on average, for masses $10<log(M/M_\odot)<11.5$ cluster galaxies appear to be smaller at fixed stellar mass than field galaxies. 
Interestingly, in the same stellar mass range but at lower redshift,  \cite{2011arXiv1109.5698C}  find exactly the opposite trend using DEEP3 \citep{2011ApJS..193...14C}. Larger sizes in the cluster environment are also observed at $z=1.62$ by \cite{2012ApJ...750...93P}  with CANDELS  data for passive galaxies with stellar masses larger than $log(M/M_\odot)\sim10.5$. 
However, other two works \citep{2010MNRAS.402..282M, 2010ApJ...709..512R} do not find any trend with environment  at $z<0.4$ and $z\sim1.2$ respectively. The differences between these works are somehow puzzling but might come from the different sample selections (e.g.: based on color or S\'ersic index vs visual morphology classification) and/or the way environment is measured (local vs. global) and/or low statistics (high redshift results indeed rely on some tens of galaxies). 

We will address here these questions by carefully selecting a statistically significant sample of galaxies from the  recently published COSMOS X-ray detected group galaxy sample \citep{2011arXiv1109.6040G, 2007ApJS..172..182F} as compared to the field. Within this sample we will dissect the properties of passive galaxies and look at the effects on the size growth of the galaxies with different morphology in different environments.


The paper is organized as follows: in section~\ref{sec:data} we describe the datasets used in this work and the derived quantities (morphologies, sizes and stellar masses) and in sections~\ref{sec:results} and~\ref{sec:disc} we present and discuss our main results.
Throughout the paper we consider a standard $\Lambda$CDM cosmology ($\Omega_M=0.3$, $\Omega_\Lambda=0.7$).

\section[]{Data and analysis}
\label{sec:data}
\subsection{Datasets}

We use two samples of galaxies from the COSMOS survey belonging to two different environments: groups and field.

The group sample is composed of groups in the COSMOS field that have been detected as extended X-ray emitters \citep{2007ApJS..172..182F}, which is a strong signature of virialized structures, and have several spectroscopic confirmed members. The full sample contains groups with halo masses from $M_{200C}/M_{\odot}\sim10^{13}$ to $\sim10^{14}$ as measured by weak-lensing \citep{2010ApJ...709...97L} and spans the redshift range $0.2<z<1.0$. Group members have been selected  based on photometric redshifts (and spectroscopic redshifts when available) derived from the extensive COSMOS multi-wavelength imaging \citep{2009ApJ...690.1236I}. 
 For this work, we use two group samples, for which details on the galaxy selection can be found in \cite{2011arXiv1109.6040G} who carried out a careful analysis of potential biases and contaminations.
 \begin{enumerate}
 \item We call the first sample {\it central} which includes only galaxies within $0.5R_{200C}$ and a probability of being a group member greater than 0.5. We expect then a contamination of $\sim15\%$ , and a completeness of $\sim90\%$ \citep{2011arXiv1109.6040G}.
 \item The second {\it larger} sample includes galaxies within within R200C and is mainly used to increase statistics, using always the central sample to control contamination effects if environment is discussed. For the larger selection the contamination is $\sim30\%$ and the completeness $\sim90\%$~\citep{2011arXiv1109.6040G}.
 \end{enumerate}  
 To both samples, we apply a magnitude cut $I814(AB) < 24$ mag required for size estimates and morphology classification as explained in sections~\ref{sec:sizes} and~\ref{sec:morpho}. \\

 


Field galaxies are selected in the COSMOS field in the same redshift range as group members and with the same magnitude cut but making sure that they do not belong to any detected group (e.g. with $GROUP_{ID}=-1$ in the \cite{2011arXiv1109.6040G} group catalog). We assume that these galaxies lie in a dark matter halo of mass $M_{200C}/M_{\odot}<10^{13.2}$, otherwise they should have been detected as groups members  -- see Fig 1 from \cite{2011arXiv1109.6040G}. The field sample contains 3760 galaxies.

\subsection{Sizes and masses}
\label{sec:sizes}
Sizes of all galaxies have been estimated on the COSMOS HST/ACS F814W images (Mosaic v2.0, \citealp{2007ApJS..172..196K}) using \textsc{galapagos} \citep{2012MNRAS.422..449B} which is an IDL based pipeline to run \textsc{Sextractor} \citep{1996A&AS..117..393B} and \textsc{Galfit} \citep{2002AJ....124..266P} together. We basically fit every galaxy in the field and in the group sample with a 2D S\'ersic profile \citep{1968adga.book.....S} using the default \textsc{galapagos} parameters as described and tested  in \cite{2007ApJS..172..615H}. {Our number of failures (i.e. fits that do not converge) is less than $5\%$. The point spread function (PSF) used for the fitting is taken from \cite{2007ApJS..172..203R} who computed spatially-varying model PSFs for the COSMOS survey taking into account variations in the effective HST focus positions. For this work we used a single PSF (estimated at the average focus position for the COSMOS survey ($\Delta=-2\mu$m)) for all the galaxies after checking that it does not introduce significant biases (see below).

The reliability of our size estimates is estimated by placing mock galaxies ($1<n<8$, $0.1^{"}<r_e<1.5^{"}$, $17<I<24$~mag) in a real background. In our simulations we also explore the possible biases due to local over densities by dropping the mock galaxies in the same positions as in real high redshift clusters. We also explore in the simulations the possible effects of a variable PSF by using a different PSF for simulating and for fitting. Both PSFs are taken from the \cite{2007ApJS..172..203R} PSFs models using the tabulation of the positional dependence of the PSF.We find that our size measurements are unbiased ($|<(r_{e,out}-r_{e,in})/r_{e,in}>|<0.1$) with a reasonable scatter ($\sqrt{Var[(r_{e,out}-r_{e,in})/r_{e,in}]}<0.2$) up to $I_{814}(AB)<24$~mag, $r_{e,in}<1.0$ arcsec and $\mu_{814W}<24$ $mag \times arcsec^{-2}$(Fig.~\ref{fig:galfit_simus} and Delaye et al., in prep). All galaxies in our sample have surface brightness brighter than $\mu_{814W}=24$ $mag \times arcsec^{-2}$ so our size estimates are reliable according to the simulations. 

Since we are using the same wavelength to estimate sizes up to $z\sim1$,  we estimate sizes in different rest-frames at different redshifts (e.g. from  the r-band rest-frame at $z\sim0.2$ to the B-band rest-frame at $z\sim1$).
Recent work by \cite{2011ApJ...739L..44D} and \cite{2011ApJ...743...96C} shows that sizes in the ultraviolet and optical rest--frame strongly correlate, with sizes in the UV rest-frame being only $\sim10\%$ smaller than those in the optical (see also \citealp{2012MNRAS.tmpL.424C}). Therefore, an additional (small) artificial size difference could be created when estimating the evolution between low and high redshift data. However, as shown in section~\ref{sec:results} and in figure~\ref{fig:passive_evol}, the size evolution we measure in our sample for early-type galaxies is fully consistent with previous published results, specially with the ones of \cite{2012ApJ...746..162N} who computed sizes in a unique rest-frame optical filter. Therefore we do not expect a significant bias due to this issue. 


Throughout the paper we will use circularized effective radii as primary size estimator, i.e. $r_e^{circ}=r_e^{fit}\times\sqrt{b/a}$.

Stellar masses are computed through spectral energy distribution (SED) fitting using the Bayesian code described in \cite{2006ApJ...651..120B} (KB06) which employs \cite{2003MNRAS.344.1000B} (BC03) models with a Chabrier IMF, and published in  \cite{2011arXiv1109.6040G} catalogs. More details can be found in section 2.3 of \cite{2010ApJ...709...97L}.

\begin{figure*}
\includegraphics[scale=0.75]{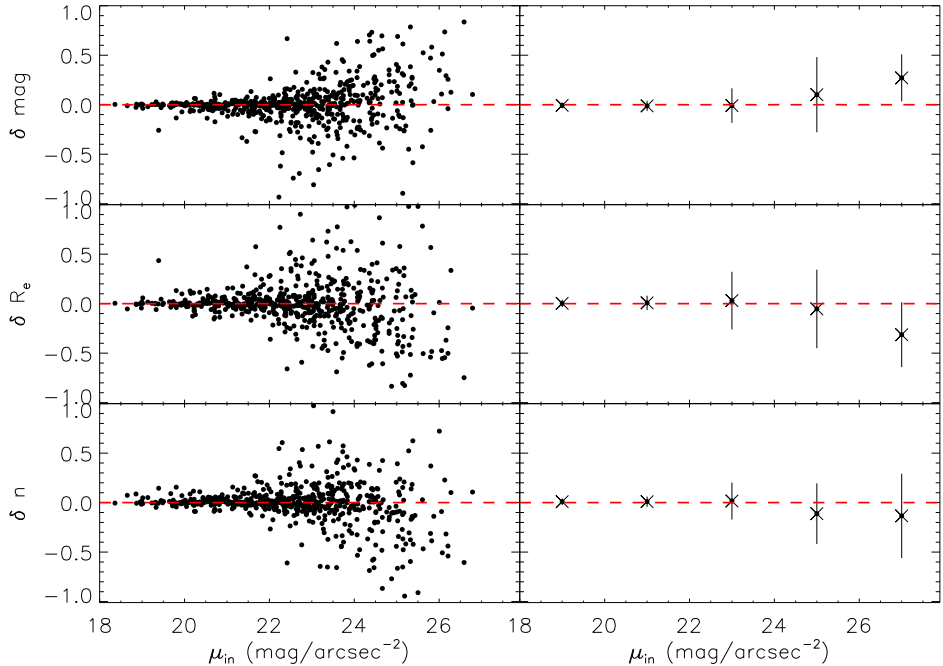}
\caption{ Results of simulations to assess the accuracy of our size estimates. Left panels show the difference between the input and output magnitudes (top left), effective radii (middle left) and S\'ersic index (bottom left) for individual simulated galaxies as a function of surface brightness. Right panels show  the average bias and dispersions for different surface brightness bins.  }
\label{fig:galfit_simus}
\end{figure*}

\subsection{Morphology}
\label{sec:morpho}

Galaxy morphologies in the group and field samples are also computed on the HST/ACS F814W images with \textsc{galSVM}, a code for automated morphological classification based on support vector machines specially designed for high redshift data \citep{2008A&A...478..971H, 2009A&A...497..743H}. The code requires a visually classified training set which is used to simulate galaxies at higher redshift. The training set used in this work is a combination of the \cite{2010ApJS..186..427N} sample of $\sim14000$ galaxies and the EFIGI project \citep{2011A&A...532A..74B} both from the Sloan Digital Sky Survey DR4. Our final training set therefore contains $\sim20.000$ galaxies.\footnote{An updated and stable version of the code as well as the training set used for this work are available at http://gepicom04.obspm.fr/galSVM/Home.html}
We follow the bayesian approach presented in \cite{2011A&A...525A.157H} to associate to every galaxy 4 probabilities of being in 4 morphological classes as defined in the local universe by \cite{2010ApJS..186..427N}, i.e. ellipticals ($E0,E+$), lenticulars ($ S0-, S0, S0+, S0/a$), early spirals ($Sa,Sab,Sb,Sbc$) or late spirals ($Sc, Sd, Sdm, Sm, Im$):  $P(E), P(S0), P(Sab), P(Scd)$. Errors in probabilities are computed by bootstrapping, i.e. we repeat the classification 10 times with randomly selected training sets from the main sample and keep the average probability for each galaxy as the final classification (see \citealp{2011A&A...525A.157H} for more details). 
We then select as elliptical and lenticular galaxies the galaxies for which $max(P(E), P(S0), P(Sab), P(Scd))=P(E)$ and $max(P(E), P(S0), P(Sab), P(Scd))=P(S0)$ respectively. Early-type galaxies are then defined as the combination of both populations. With this selection, the completeness for early-type galaxies is $\sim95\%$ and the contamination rate is expected to be less than 7\% as stated by detailed comparisons with visual morphological classifications at $z\sim1.3$ from \citep{2009ApJ...690...42M, 2012arXiv1205.1785M} in the Lynx super cluster at  $z=1.26$. Our classification is therefore very close to a visual classification. This point is also confirmed by the fact that the axis ratio distribution of our early-type sample is very close to the one reported by \cite{2011arXiv1111.6993B} from a visually classified sample (see Fig.~\ref{fig:ells_sos} and their figure 2).

To compare our morphological classification with those often used in the literature, we estimated the contamination that would suffer a selection of early-type galaxies, based on a simple $n>2.5$ cut. We find that such a sample would be contaminated by approximately $50\%$ of early spirals (see also \citealp{2012arXiv1205.1785M}). We refer to section~\ref{sec:results} for a detailed analysis of the implications of such a selection.

Separating ellipticals from lenticulars is extremely challenging even by eye, so our classification is necessarily more contaminated. Simulations show that we are close to $30\%$ uncertainties, similar to those observed in other works (e.g. by visual morphological classifications by \citealp{2005ApJ...623..721P}). Our visually trained probabilistic approach for galaxy classification allows us to distinguish galaxies based on a quantitative separation in the parameter space that corresponds to that usually used to separate S0s and ellipticals in the local Universe (e.g.  by visual classification). Indeed, the stellar mass and axis ratios distributions of the two populations shown in Fig.~\ref{fig:ells_sos} present different behaviors, as stated by Kuiper tests ($P<0.6$), confirming that we are seeing separate populations and that the classification is not random. Ellipticals appear rounder, more massive on average, with slightly higher S\'ersic indices and larger sizes. In order to double check our separation between lenticulars and ellipticals we performed two additional tests. 
First, we took the visually classified sample of \cite{2010ApJS..186..427N} in the SDSS, cross--correlated it with the S\'ersic decompositions recently published by \cite{2011ApJS..196...11S} and looked at the same properties we studied for our high redshift sample. Results are shown in Fig.~\ref{fig:props_SDSS}. We do clearly observe the same trends, i.e. lenticulars are also more elongated and slightly more compact than ellipticals, even though it is less pronounced than in the high redshift sample. This test confirms that our E/S0 classification is consistent with the local visual classification, as expected.
As a second step, two of us (MHC and SM) made visual classifications of our two automatically  defined classes. We find that the two classification of E/S0s agree at $\sim80\%$ at $z<0.5$ and $\sim70\%$ at $z>0.5$ which is fully consistent with the results from simulations and also with the differences found between independent human classifiers (e.g. \citealp{2005ApJ...623..721P}). However, in order to make sure that our results are not biased because of the automated method used, we double check our main results using visual classifications (see appendix~\ref{app:morpho}) and comment in our analysis in subsequent sections whenever there is a significant difference.

We deduce that the population of elliptical galaxies selected through our method is dominated by pure bulge systems whereas an important fraction of what we call lenticulars have a disk component (but not observed arms) even if they are still bulge dominated. We also notice that a simple S\'ersic based selection of any kind would not allow the separation between these two populations.

\begin{figure*}
\includegraphics[scale=0.75]{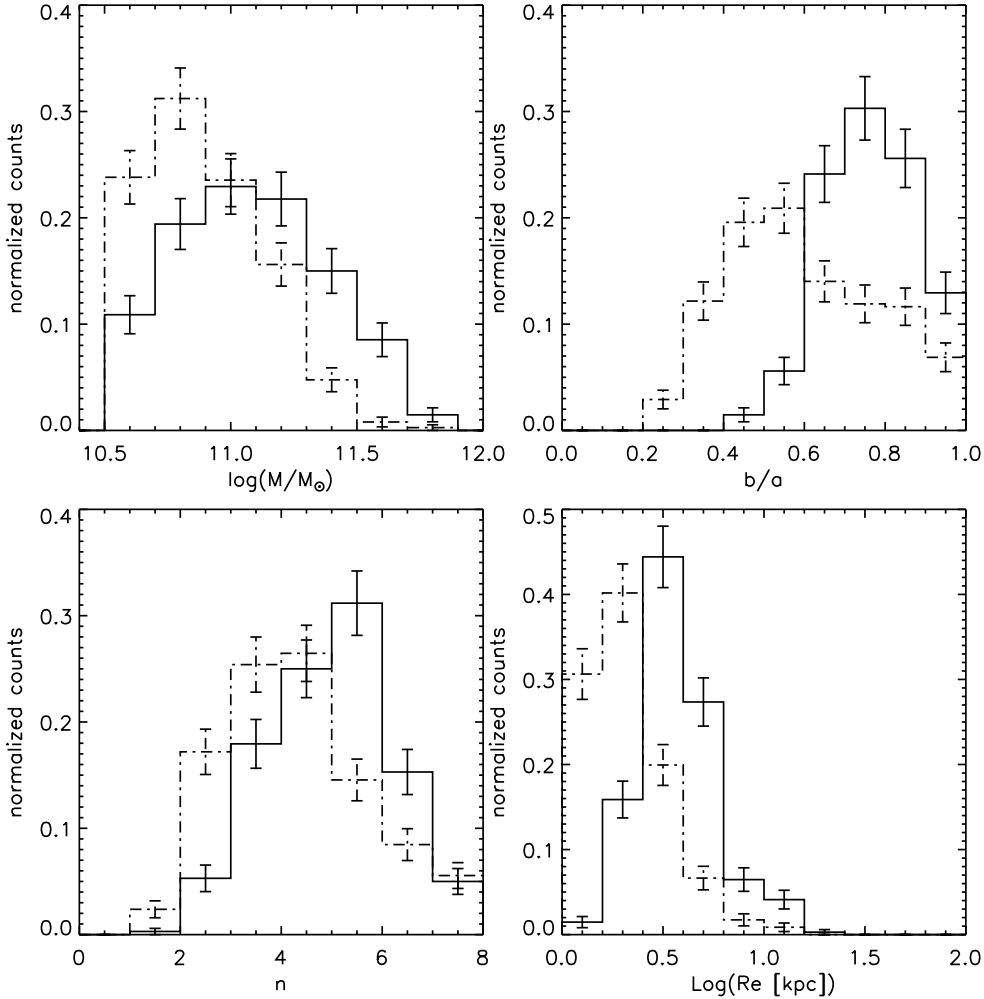}
\caption{ Stellar mass (top left panel), axis ratio (top right panel), S\'ersic index (bottom left panel) and size distributions (bottom right panel) of morphologically classified ellipticals (solid line) and S0s (dashed line) with the automated procedure described in the text. Error bars are Poissonian. S0s and ellipticals have different parameter distributions, which proves that the algorithm indeed separates two different populations of galaxies and does it in a quantitative and reproducible way.}
\label{fig:ells_sos}
\end{figure*}

\begin{figure*}
\includegraphics[scale=0.75]{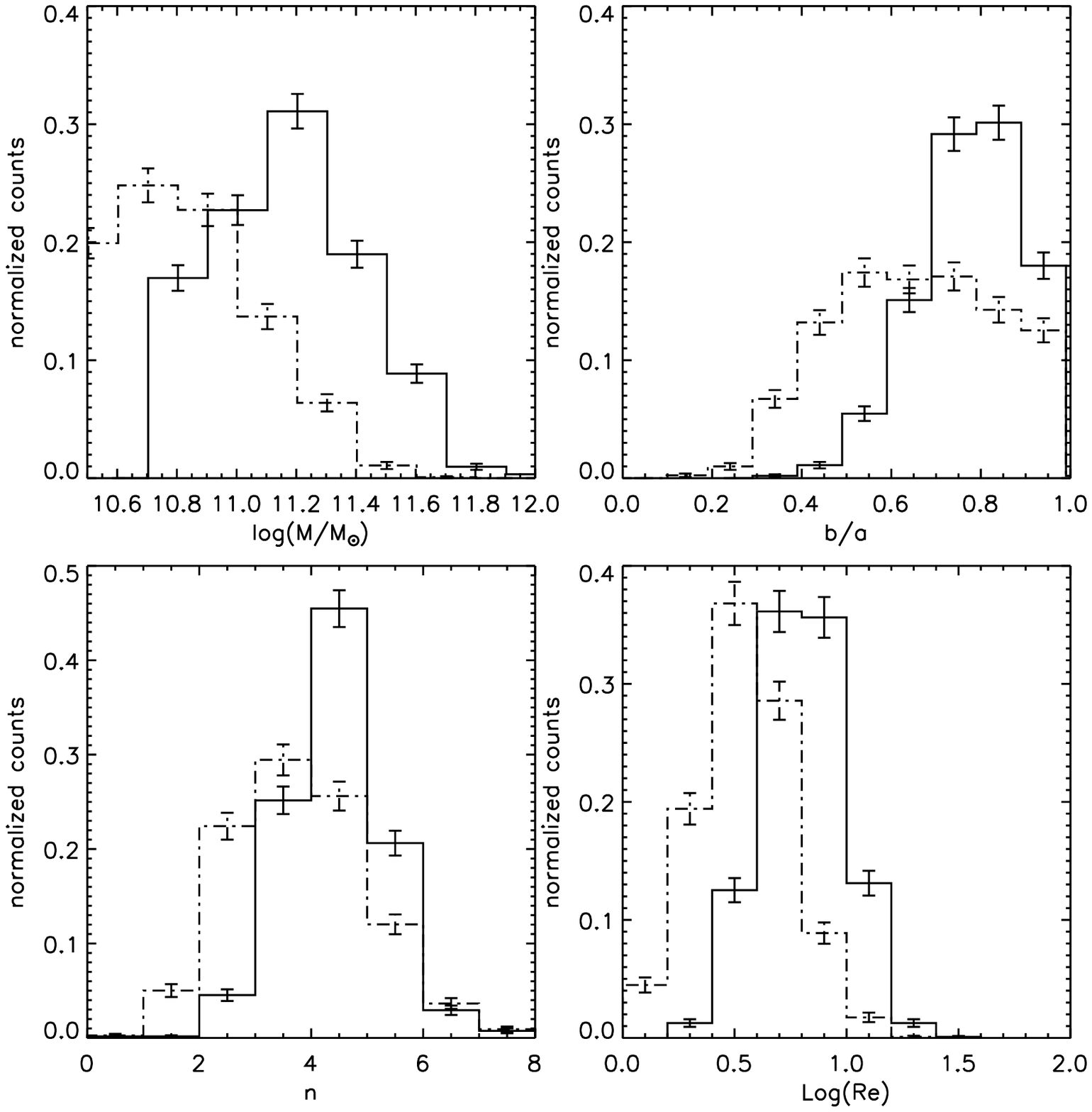}
\caption{ Same plot as figure~\ref{fig:ells_sos} but for a local sample from the SDSS (Simard et al. 2011). The same trends as for the high redshift sample are observed. }
\label{fig:props_SDSS}
\end{figure*}

\subsection{Completeness}
\label{sec:compl}

Completeness of our sample is mainly driven by the apparent magnitude cut ($I_{814}(AB)<24$~mag) required to properly estimate morphologies and sizes. Our main results in the following are shown as a function of stellar mass. Therefore it is important to understand how this magnitude selection is translated in terms of stellar mass completeness to estimate how much the results might be affected by selection effects. Since in this work we focus on passive ETGs (see section~\ref{sec:sel}), all the completeness values are given for that population.

We used an approach similar to \citealp{2010A&A...523A..13P} and \citealp{2012A&A...538A.104G}. For each passive galaxy we compute its limiting stellar mass ($M_*^{lim}$) which is the stellar mass it would have if its apparent magnitude was equal to the limiting magnitude of the survey ($I_{814}(AB)=24$~mag in our case). This value is given by the relation $Log(M_*^{lim})=Log(M_*)-0.4(I-I_{lim})$ following Pozzetti et al. (2010). We compute this limiting mass for the $20\%$ faintest galaxies in each redshift bin and estimate the 95\% completeness as the 95th percentile of the resulting distribution. Following this approach we obtain a 95\% completeness for galaxies with stellar masses greater than $log(M_*/M_\odot)\sim9.56$ at $z\sim0.2$ and $log(M_*/M_\odot)\sim10.57$ at $z\sim1$ (fig.~\ref{fig:compl_pozz}). 

We notice that the mass completeness for the COSMOS sample has been largely discussed in the literature (e.g. \citealp{2009A&A...503..379T}, \citealp{2012A&A...538A.104G}, \citealp{2010A&A...523A..13P}, \citealp{2009A&A...505..463M}) finding similar values. 

In order to further check that low surface brightness objects are not lost, we carried a new set of simulations close to the ones performed to calibrate the size recovery. We modeled 2000 ETGs ($B/T>0.6$) with effective radii varying from $0.1$ to $1.5$ arcsec and magnitudes at the faint end of the survey detection limit ($23<I_{814}<26$) , dropped them in real background and computed the detection rate as a function of magnitude and size (fig~\ref{fig:compl_simu}). Galaxies with $I_{814}<24$ of all sizes (and hence all surface brightness) are all detected thus confirming that our analysis sample is complete .

\begin{figure}
\includegraphics[scale=0.75]{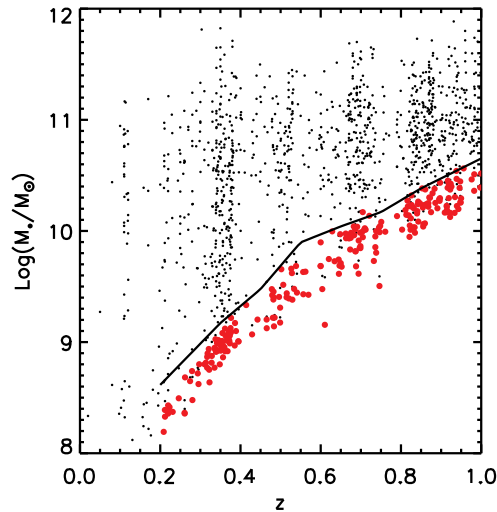}
\caption{ Stellar mass as a function of redshift of passive ETGs.  Red filled big circles are $M_*^{lim}$ and the black line shows the $95\%$ completeness level. See text and Pozzetti et al. (2010) for more details.}
\label{fig:compl_pozz}
\end{figure}

\begin{figure}
\includegraphics[scale=0.75]{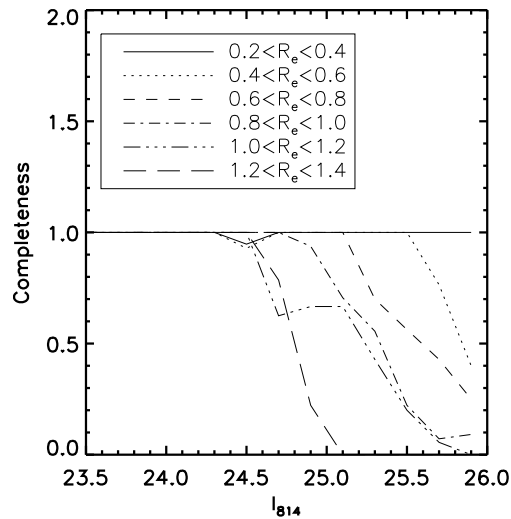}
\caption{ Completeness of our sample as a function of magnitude and size in arcsec (see text for details). Objects with $F814W<24$ are all detected. }
\label{fig:compl_simu}
\end{figure}

\section{Results}
\label{sec:results}

\subsection{ Sample selection}
\label{sec:sel}

We investigate in this section how the sample selection affects the observed mass-size relation and size evolution of selected galaxies. We compare first three selections usually found in other works: 
\begin{enumerate}
\item a passive selection. This selection includes all quiescent galaxies with $log(M/M_\odot)>10.5$. The mass cut ensures that we are selecting a volume limited sample without being affected by incompleteness (see section~\ref{sec:compl}). Throughout this work, we will select quiescent or passive galaxies based on the M(NUV)-M(R) dust corrected rest--frame color as computed by \cite{2009ApJ...690.1236I} by means of SED fitting. \cite{2009ApJ...690.1236I} have shown that a $M(NUV)-M(R)>3.5$ selection results in a good separation of passive and star-forming galaxies. See also \cite{2012A&A...538A.104G} for a discussion on this selection on the same sample as the one used for this work. 
\item a S\'ersic based selection, i.e. we apply a simple $n>2.5$ cut as usually done as well as the same stellar mass selection ($log(M/M_\odot)>10.5$).
\item an ETG selection in which we select all early type galaxies (ellipticals and lenticulars) with $log(M/M_\odot)>10.5$ from our morphological classification detailed in section~\ref{sec:morpho} independently of the star formation activity.
\end{enumerate}

\begin{figure*}
\includegraphics[scale=0.7]{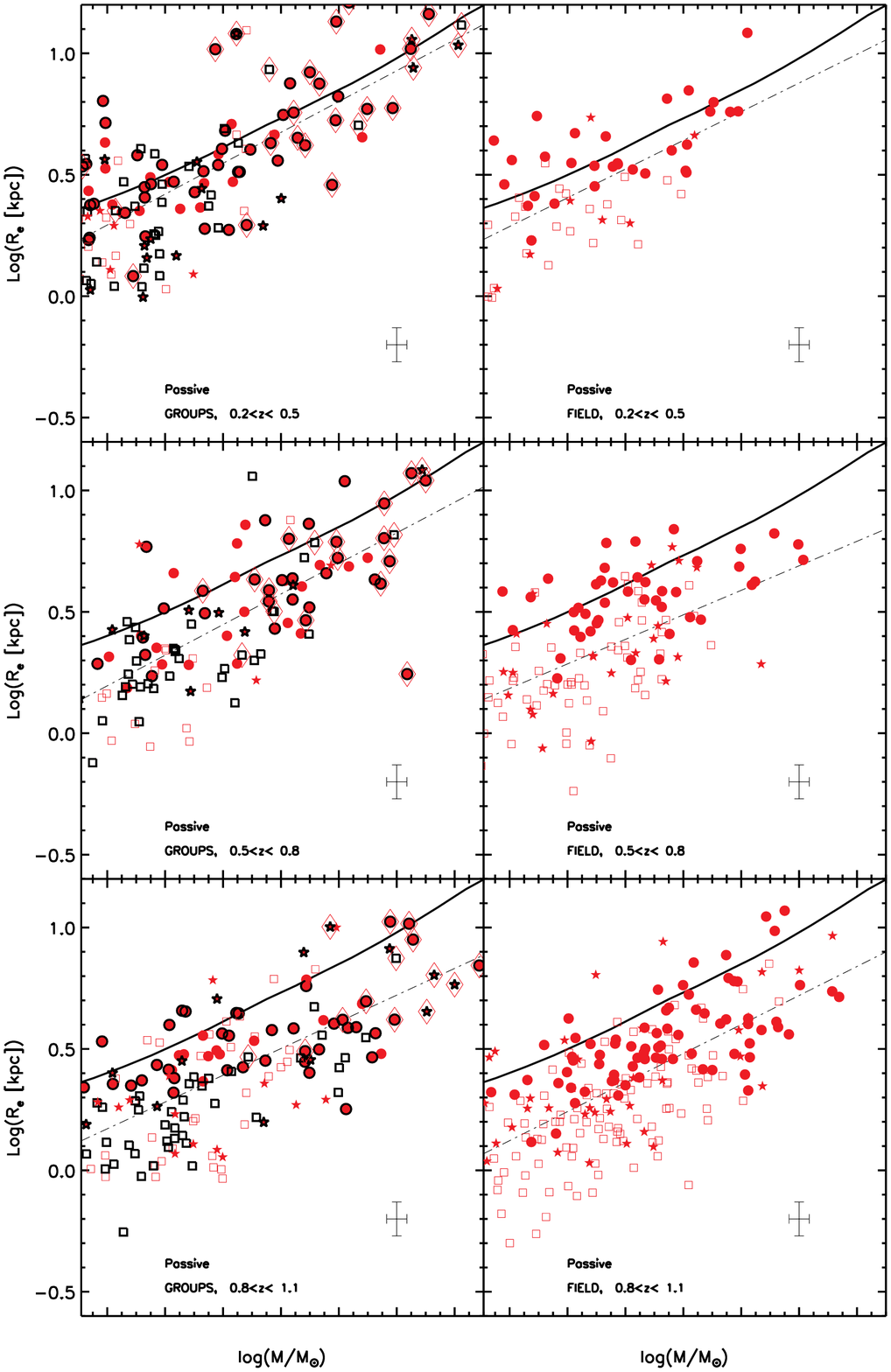}
\caption{Mass-size relation of passive galaxies in groups (left column) and field environments (right column). Filled circles are ellipticals, empty squares are lenticulars and stars are late-type galaxies (including early and late-type spirals). Symbols with a black thick contour show \emph{central} group members ($r<0.5R_{200C}$ and $P_{MEM}>0.5$). Diamonds show the position of BGGs. We also show on the bottom right corner the typical error bars for sizes and stellar masses. The solid line shows the local relation as measured by Bernardi et al. (2010) and dashed-dotted lines are the best fits to the whole passive population. The distribution in the mass-size plane of the different morphological types are not the same. A pure \emph{star-formation} selection does not ensure that we are studying just bulge growth but also galaxies with a disk component which might follow a different evolutionary path. }
\label{fig:mass_size_passive}
\end{figure*}

\begin{figure*}
\includegraphics[scale=0.7]{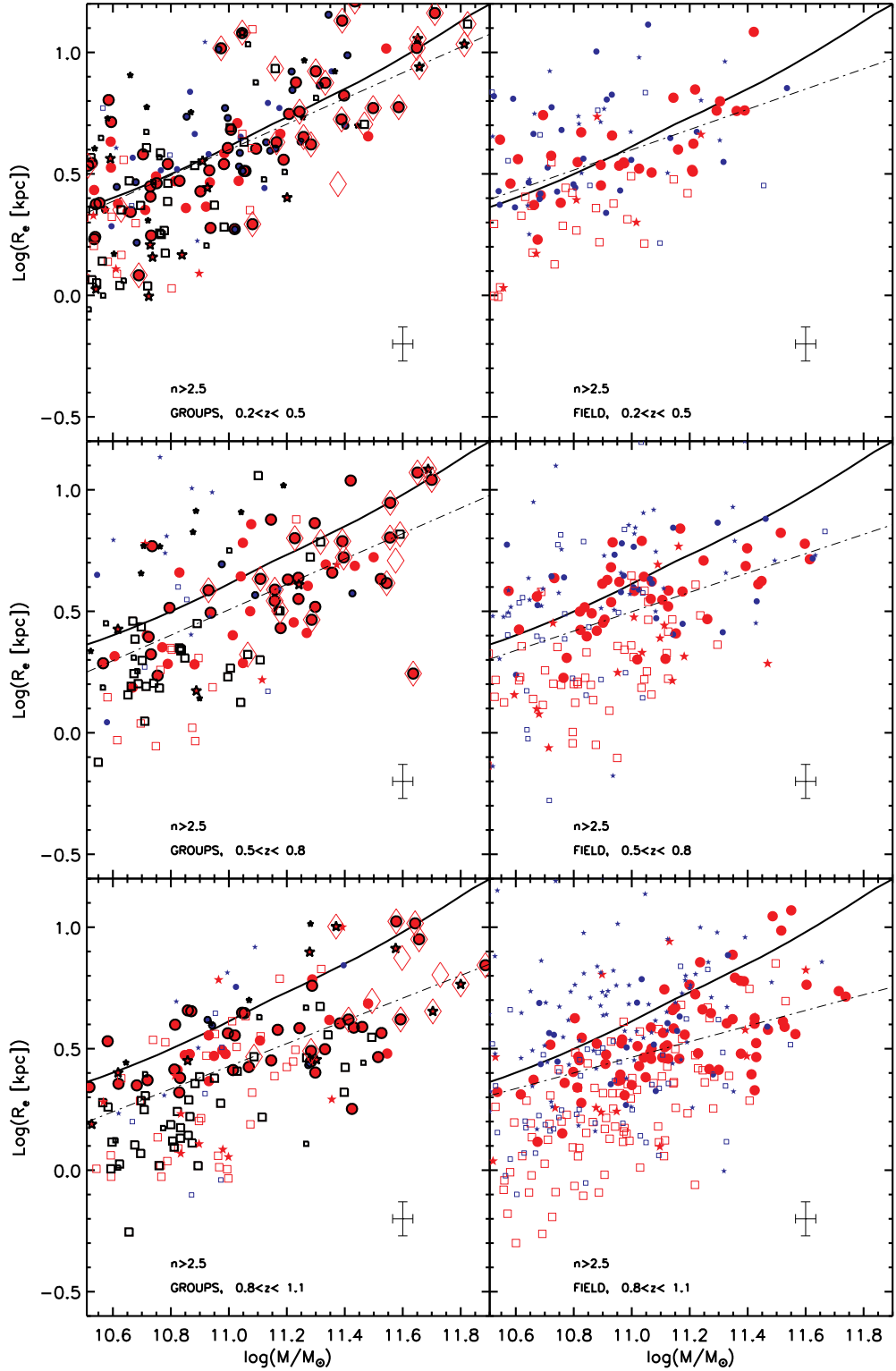}
\caption{Mass-size relation of $n>2.5$ galaxies in groups (left column) and field environments (right column). Filled circles are ellipticals, empty squares are lenticulars and stars are late-type galaxies (including early and late-type spirals). Symbols with a black thick contour show \emph{central} group members ($r<0.5R_{200C}$ and $P_{MEM}>0.5$). Diamonds show the position of BGGs. Blue small /red big symbols show active/passive galaxies respectively. We also show on the bottom right corner the typical error bars for sizes and stellar masses. The solid line shows the local relation as measured by Bernardi et al. (2010) and dashed-dotted lines are the best fits to the whole $n>2.5$ population.  A S\'ersic index based selection contains an important fraction of star-forming galaxies which tend to increase the scatter of the relation. The effect seems to be more pronounced in the field population so it could yield to observed differences in the evolution of the mass-size relation with environment } 
\label{fig:mass_size_sersic}
\end{figure*}

\begin{figure*}
\includegraphics[scale=0.7]{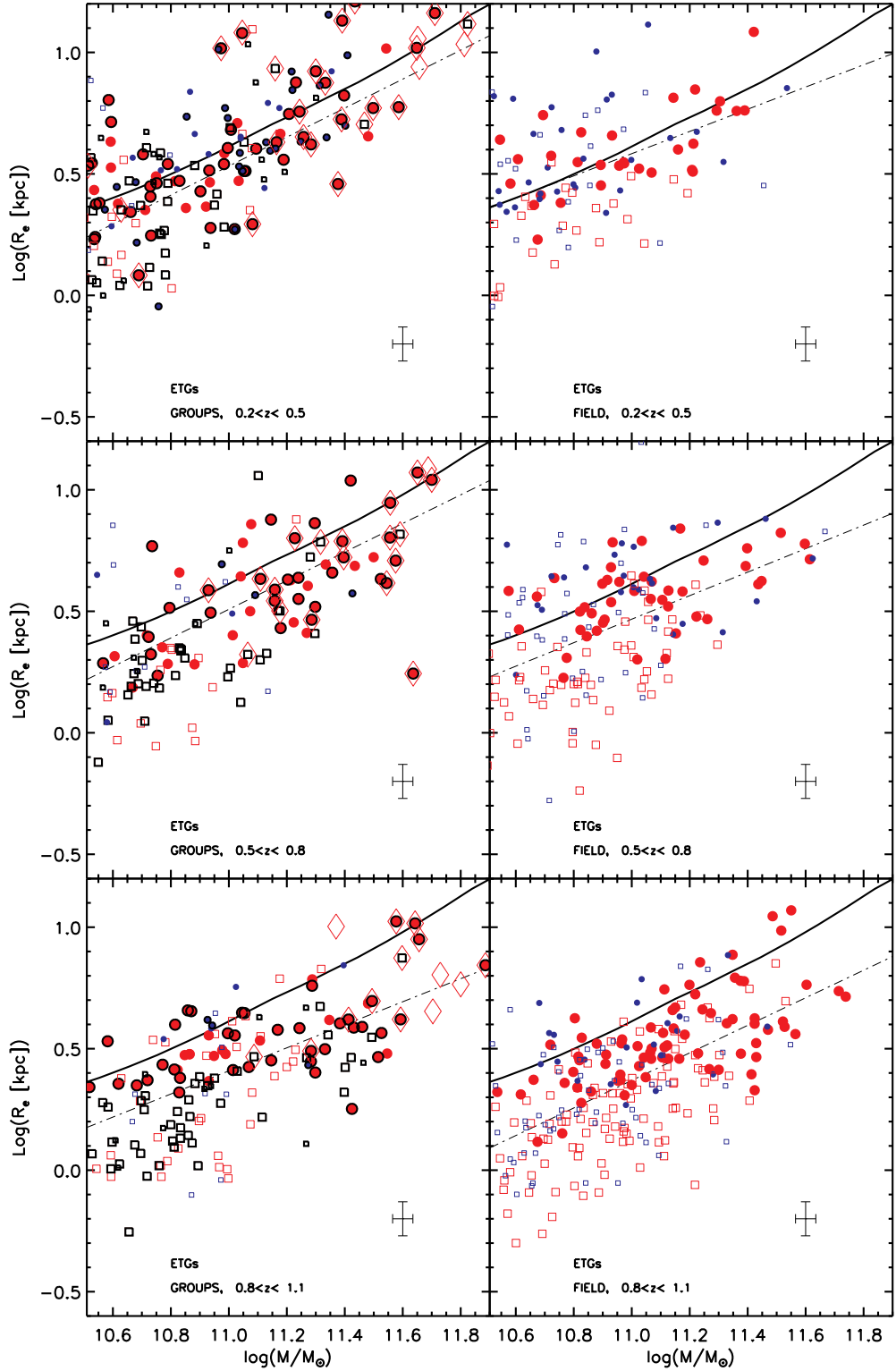}
\caption{Mass-size relation of early-type galaxies in groups (left column) and field environments (right column). Filled circles are ellipticals and empty squares are lenticulars. Symbols with a black thick contour show \emph{central} group members ($r<0.5R_{200C}$ and $P_{MEM}>0.5$). Diamonds show the position of BGGs. Blue small/red big symbols show active/passive galaxies respectively. We also show on the bottom right corner the typical error bars for sizes and stellar masses. The solid line shows the local relation as measured by Bernardi et al. (2010) and dashed-dotted lines are the best fits to the whole early-type population.} 
\label{fig:mass_size_morpho}
\end{figure*}

\begin{figure*}
\includegraphics{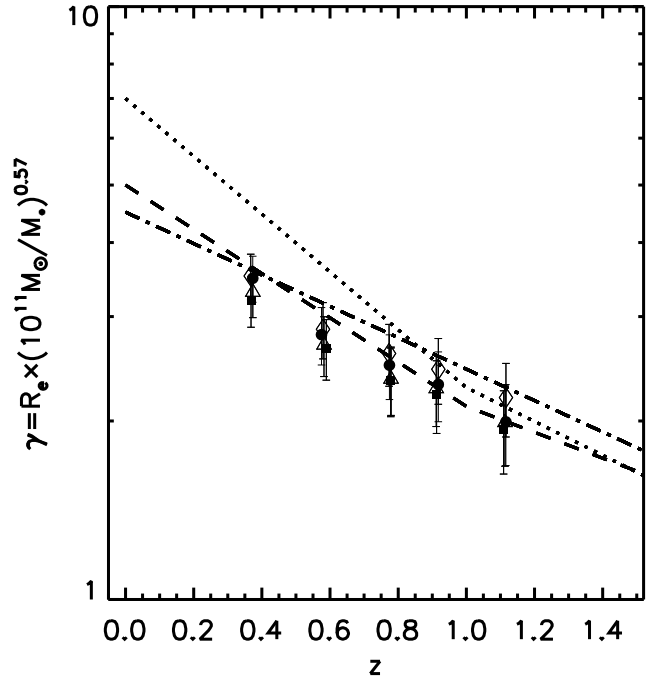}
\caption{Mass-normalized radius for our sample for galaxies with masses $log(M/M_\odot)>10.7$,  as a function of redshift for different selections compared to recent published results. Circles are ETGs, squares passive galaxies, triangles passive ETGs and diamonds $n>2.5$ galaxies. Dashed line shows Cimatti et al. (2012) fit, dotted-dashed line is Newman et al. (2012) and dotted line is Damjanov et al. (2011). Samples selected using the S\'ersic index tend to show larger sizes, because of the contamination from passive spirals, which is larger in field samples. However, these differences are within 1$\sigma$. }
\label{fig:passive_evol}
\end{figure*}

\begin{figure*}
\includegraphics{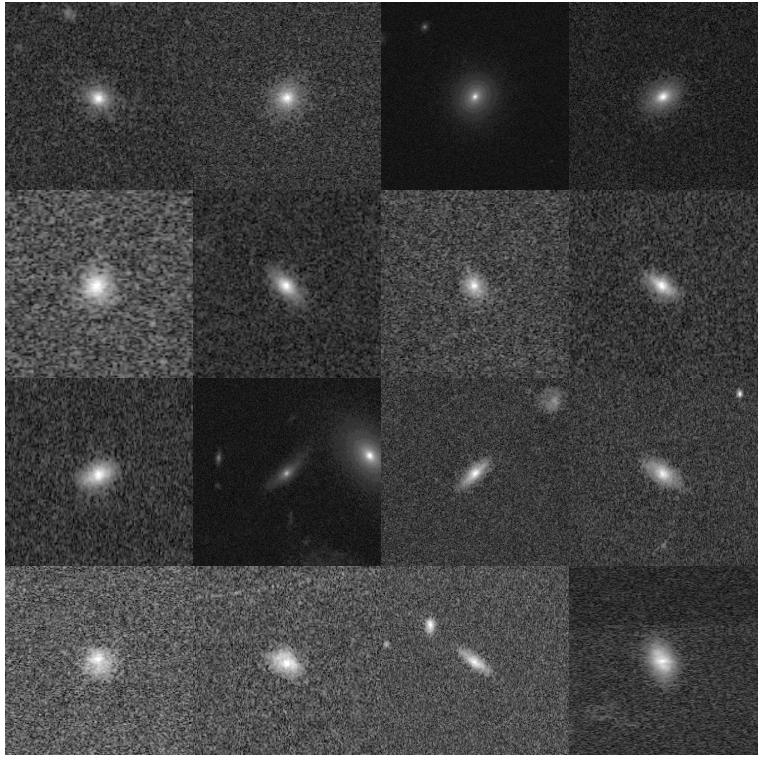}
\caption{Example stamps of passive galaxies of 4 morphological classes. First row: ellipticals, second row: S0s, third row: early-spirals and last row: late-spirals. Stamp sizes are: $1^{"}\times1^{"}$. }
\label{fig:morpho_stamps}
\end{figure*}

\begin{figure*}
\includegraphics[scale=0.75]{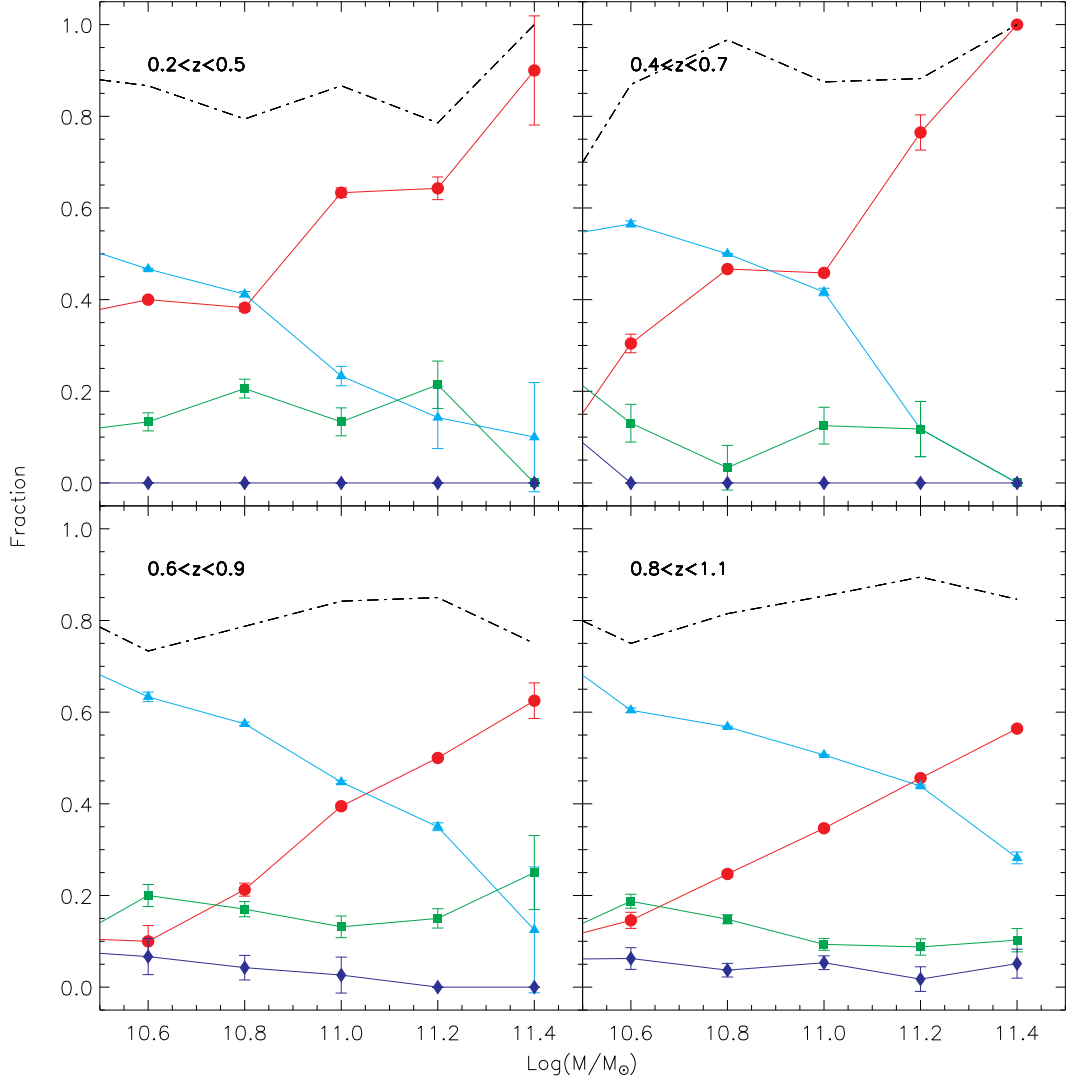}
\caption{Morphological fractions of passive galaxies as a function of stellar mass and redshift. Red circles: ellipticals, cyan triangles: S0s, green squares: early-spirals, blue diamonds: late-spirals. The black dashed line shows the fraction of early-type galaxies (ellipticals and S0s together). The uncertainties are calculated following Gehrels (1986; see Section 3 for binomial statistics; see also Mei et
al. 2009). These approximations apply even when ratios of different events are calculated from small numbers, and yield the lower and upper limits of a binomial distribution within the 84\% confidence limit, corresponding to 1$\sigma$. Note that using this conservative approach our uncertainties are slightly overestimated (Cameron et al. 2011). }
\label{fig:morpho_fractions}
\end{figure*}

\begin{table*}
\begin{tabular}{cccccc}
\hline\hline\noalign{\smallskip}
Redshift & Environment & Sample & a & log(b) & N\\ 
\hline\hline\noalign{\smallskip}
$0.2<z<0.5$ & group & Pall & $0.52\pm0.03$ & $-5.25\pm0.34$ & 128\\
 &  & PEs & $0.47\pm0.04$ & $-4.65\pm0.49$ & 65\\
  &  & PS0s & $0.49\pm0.05$ & $-4.91\pm0.57$ & 43\\
    &  & $n>2.5$ & $0.52\pm0.05$ & $-5.21\pm0.37$ & 263\\
      &  & ETGs & $0.59\pm0.06$ & $-6.02\pm0.66$ & 133\\
  & field & Pall & $0.59\pm0.09$ & $-5.9\pm0.99$ & 59 \\
  &  & PEs & $0.41\pm0.04$ & $-4.65\pm0.49$ & 30\\
    &  & PS0s & $0.49\pm0.05$ & $-4.91\pm0.57$ & 22\\
    &  &  $n>2.5$ & $0.41\pm0.05$ & $-3.98\pm0.56$ & 161\\
     &  & ETGs & $0.45\pm0.08$ & $-4.44\pm0.92$ & 99\\
  $0.5<z<0.8$ & group & Pall & $0.56\pm0.04$ & $5.27\pm0.45$ & 110\\
   &  & PEs & $0.40\pm0.07$ & $-3.85\pm0.8$ & 52\\
      &  & PS0s & $0.51\pm0.06$ & $-5.27\pm0.72$ & 46\\
      &  & $n>2.5$ & $0.58\pm0.05$ & $-5.24\pm0.52$ & 176\\
      &  & ETGs & $0.58\pm0.06$ & $-5.97\pm0.66$ & 155\\
       & field & Pall & $0.50\pm0.11$ & $-5.16\pm1.27$ & 123\\
        &  & PEs & $0.27\pm0.07$ & $-3.85\pm0.8$ & 43\\
          &  & PS0s & $0.51\pm0.06$ & $-5.27\pm0.72$ & 54\\
          &  & $n>2.5$ & $0.39\pm0.04$ & $-3.88\pm0.49$ & 250\\
           &  & ETGs & $0.48\pm0.07$ & $-4.85\pm0.78$ & 158\\
          
           $0.8<z<1.0$ & group & Pall & $0.49\pm0.04$ & $-4.98\pm0.41$ & 155\\
   &  & PEs & $0.35\pm0.05$ & $-3.38\pm0.60$ & 53\\
      &  & PS0s & $0.42\pm0.05$ & $-4.37\pm0.57$ & 73\\
      &  & $n>2.5$ & $0.46\pm0.04$ & $-4.71\pm0.46$ & 209\\
      &  & ETGs & $0.47\pm0.05$ & $-4.81\pm0.59$ & 148\\
       & field & Pall & $0.59\pm0.05$ & $-6.21\pm0.52$ & 210\\
        &  & PEs & $0.43\pm0.05$ & $-3.38\pm0.60$ & 65\\
          &  & PS0s & $0.42\pm0.05$ & $-4.37\pm0.57$ & 99\\
          &  & $n>2.5$ & $0.32\pm0.03$ & $-3.07\pm0.41$ & 394\\
           &  & ETGs & $0.56\pm0.04$ & $-5.86\pm0.51$ & 252\\

\hline

\end{tabular}
\caption{Power-law fitting parameters to the mass-size relation ($r_e=(\frac{M}{M\odot})^a+b$) for different galaxy selections and environments. Pall stands for \emph{Passive all} (sample (i) from the text), PEs is \emph{Passive ellipticals}, PS0s means \emph{Passive lenticulars}, $n>2.5$ are galaxies from a S\'ersic based selection (sample (ii)) and ETGs stands for early-type galaxies independently of star formation (sample (iii)). N is the number of objects in each bin with $log(M_*/M_\odot)>10.5$.}.
\label{tbl:mass_evol}
\end{table*}

We show in Figs.~\ref{fig:mass_size_passive} to~\ref{fig:mass_size_morpho} the mass size relations for samples (i), (ii) and (iii) respectively in different environments and redshifts and we summarize the best fit parameters computed through a standard chi square minimization in table~\ref{tbl:mass_evol}. Also are shown in the figures, with different symbols the morphologies of the objects belonging to the given selection as well as the star formation activity (quiescent or active) color coded for selection (ii) and (iii). We also show the position of the brightest group galaxies (BGGs), defined as the most massive galaxies within an NFW scale radius of the X-ray position, from the \cite{2011arXiv1109.6040G} catalog. Notice that some of them are morphologically classified as early-spirals. This is probably because of the extended halo usually found in these objects which could be interpreted as a disk component by the automated algorithms.  

Despite the fact that we see some differences between the different selections, the fit parameters are roughly consistent within $1\sigma$. Since a S\'ersic index based selection contains an important fraction of star-forming galaxies ($\sim 50\%$) (a population which presents, on average, higher sizes than ETGs; e.g. Mei et al. 2012), this population tends to increase the scatter of the relation.

A similar behavior is observed if we select galaxies just based on the morphology (ETGs, fig.~\ref{fig:mass_size_morpho}). The scatter of the relation might be increased by the presence of a star-forming population of galaxies with early-type morphology. Even if the number densities of these objects are small in the local universe (e.g \citealp{2009AJ....138..579K}) they tend to increase at high z (e.g \citealp{2010A&A...515A...3H}) and might have a consequence on the size measurements. 

We quantify the possible effects of galaxy selection in Fig.~\ref{fig:passive_evol} where we plot the mass-normalized radius ($R_e\times(10^{11}M_\odot/M_*)$) evolution for different galaxy selections (ETG, passive ETGs, $n>2.5$, passive galaxies) with $log(M/M_\odot)>10.7$ (the mass selection for this figure is chosen to match the one from \citealp{2012ApJ...746..162N}). We observe  that surprisingly, the differences between the different selection in terms of size evolution are negligible and also consistent with recent results (i.e. \citealp{2012MNRAS.tmpL.424C, 2011ApJ...739L..44D, 2012ApJ...746..162N}). We can conclude that these different selections usually found in the literature lead to similar results.

In this work we have an additional ingredient though, usually lacking in previous published results which is the morphological dissection of passive galaxies as explained in section~\ref{sec:morpho}. As a matter of fact, the population of passive galaxies is not a homogeneous population of objects. We show indeed some example stamps of massive ($log(M/M_\odot)>10.5$) passive galaxies with different morphologies in Fig.~\ref{fig:morpho_stamps}. It is easy to notice by simple visual inspection that not all passive galaxies are bulge dominated. The relative abundance of each morphological type as a function of stellar mass is quantified in Fig.~\ref{fig:morpho_fractions}, for all our passive galaxy sample. 

As expected, early-type morphologies dominate the population of passive galaxies up to $z\sim1$ at all stellar masses, being about 80\% of the total population at all redshifts. The remaining 20\% is populated by early-type spirals while late-type spiral fractions are negligible. At all redshifts up to $z\sim1$, elliptical galaxies dominate the ETG population at masses $log(M/M_\odot)>11-11.2$. At $z<0.5$, galaxies with masses $log(M/M_\odot)<11$ are around half ellipticals and half lenticular, while lenticulars dominate the low--mass fractions at $z>0.5$. In other words, an important fraction of passive galaxies from $z\sim1$ (if not the majority) could have a disk component (see also \citealp{2010ApJ...719.1969B}, \citealp{2012arXiv1205.1785M}, \citealp{2011ApJ...730...38V}). We also show in appendix, a version of figure~\ref{fig:morpho_fractions} obtained with the two independent visual classifications performed in this work. One of the classifiers (SM) finds visually more Sas than S0s with respect to the automated classification, specially at high redshift. However the separation between galaxies with disk (Sas and S0s) and without disk (ellipticals) is very similar to what is obtained with the automated classification, except that elliptical galaxies are not dominant at the high mass end in the highest redshift bin, which is also found by the second classifier (MHC).

Studying the size evolution of {\it passive galaxies all together} (as often done in several works), mixes not only ETGs with early-spirals (Sa-b), but also ETG different morphological populations (e.g. ellipticals and lenticulars)  for which the evolution is not necessarily of the same nature. 

Differences in the mass-size relation of ellipticals and lenticulars are visible in Fig.~\ref{fig:mass_size_passive} to  \ref{fig:mass_size_morpho}  and Table~\ref{tbl:mass_evol}. S0 galaxies tend to be not only less massive but also more compact with respect to ellipticals.

In the next sections, we will focus on the sample of passive galaxies (sample (i)) and will quantify these differences by studying the dependence of the size evolution of passive galaxies on different morphological types (namely ellipticals and lenticulars). The final sample contains 404  group (232 within $0.5\times R_{200C}$ and with $P_{MEM}>0.5$) and 459 field galaxies.

\subsection{Size growth of passive galaxies with different morphologies from $z\sim1$}
\label{sec:growth}

Throughout this section we normalize sizes using the local reference derived by \cite{2010MNRAS.404.2087B} on the SDSS using a clean sample of elliptical galaxies (excluding S0s). We payed special attention in properly calibrating the local reference to be as consistent as possible with our high redshift measurements. We checked for that purpose that our low redshift ($z<0.3$) data are consistent at  $1\sigma$ with their local mass-size relation which confirms that both relations are well calibrated. Sizes in each redshift bin are computed by fitting a 1-D gaussian function to the size ratio distributions and keeping the position of the peak as the adopted value. Results remain however unchanged if we use instead a classical median or a three sigma-clipped average. Uncertainties are computed through bootstrapping, i.e. we repeat the computation of each value 1000 times removing one element each time and compute the error as the $1-\sigma$ error of all the measurements. \\

 In this section, since we are mainly interested in the effects of different morphologies, we mix group (larger selection) and field galaxies in a single population to improve statistics. Both populations will be considered separately in section~\ref{sec:environment} in which we study the effects of environment in the mass--size relation.


\subsubsection{Ellipticals}
\label{sec:ells}

In fig.~\ref{fig:evol_obser}, we show the size evolution of ellipticals as compared to lenticulars in two mass bins ($10.5<Log(M/M_\odot)<11$ and $11.0<Log(M/M_\odot)<11.5$). The upper limit is chosen because there are almost no S0s with masses greater then $10^{11.5}$ solar masses (see fig.~\ref{fig:ells_sos}) and we want to compare the evolution of the two morphological types in the same mass ranges. 

Ellipticals in the mass range $11.0<Log(M/M_\odot)<11.5$ experienced a $\sim40\%\pm10\%$ size growth from $z\sim1$ to present. We find an $\alpha$ value of $\alpha=0.8\pm0.28$ when fitting the evolution with a power-law ($r_e\propto (1+z)^{-\alpha}$).  However, the most relevant feature is that, as shown in the left panel of fig.~\ref{fig:evol_obser}, below $\sim 10^{11}$ solar masses, elliptical galaxies do not experience a significant size growth, e.g their size does not evolve in a significant way with respect to local ellipticals from the SDSS ($r_e\propto (1+z)^{-0.34\pm0.17}$) (see also \citealp{2012ApJ...745..130R} for similar results at higher redshift). The mass dependence is even more clear in figure~\ref{fig:evol_ETG} in which we show in the left column, the observed size evolution for ellipticals in three bins of increasing stellar mass ($10.5<log(M/M_\odot)<11$, $11<log(M/M_\odot)<11.2$ and $11.2<log(M/M_\odot)<12$). The derived $\alpha$ value parameterizing the evolution is $\alpha=0.34\pm0.17$, $\alpha=0.63\pm0.18$ and $\alpha=0.98\pm0.18$ from low to high mass, showing a clear increasing trend with stellar mass. In appendix~\ref{app:morpho}, we show that when using visual morphological classifications the evolution in the low mass bin is stronger ($\sim30\%$) making the difference between the two first mass bins less pronounced. The trend is still preserved though (see table~\ref{tbl:alpha_values}), specially between the two last bins. One possible source of error though is that the stellar mass bins used are comparable to the expected error of the stellar mass (i.e. 0.2 dex), so the different behaviors found can be affected by contaminations of objects with lower/higher stellar masses.  We have run Monte Carlo simulations to check if the trends found are preserved with typical errors expected on the stellar mass. For that purpose, we added to every stellar mass a random shift within $3\sigma$ of the expected error in stellar mass and recomputed the median sizes 1000 times. The values found are consistent at $1\sigma$ level so we do not expect a significant contribution of this effect in our measurements.


Interestingly, this mass dependence is less pronounced when studying all ETGs or passive galaxies as a whole (right column of fig.~\ref{fig:evol_ETG}). The best fit relations for this whole sample are $r_e\propto (1+z)^{-1.01\pm0.23}$ for ETGs with stellar masses $10.5<Log(M/M_\odot)<11$ and $r_e\propto (1+z)^{-1.21\pm0.22}$  and $r_e\propto (1+z)^{-1.19\pm0.18}$ for $11<Log(M/M_\odot)<11.2$ and $11.2<Log(M/M_\odot)<12$ respectively which are fully consistent within $1\sigma$. 
 
 The mass dependence has been discussed in the recent literature providing different results. \cite{2010ApJ...713..738W} and \cite{2012ApJ...749...53R} measured a mass dependence similar to the one reported here for ellipticals, i.e. with the evolution being stronger at higher stellar masses, while other works like \cite{2011ApJ...739L..44D} suggest the slope of the mass-size relation is mass independent. We show here that the results might depend on how the selection is performed. When selecting pure passive bulges a mass dependent evolution seems to emerge. As a matter of fact,  \cite{2010ApJ...713..738W} select passive galaxies based on the specific star formation rate (instead of the red sequence for other works). They might be removing from their sample more \emph{disky} galaxies (also removed from our elliptical sample and not from an ETG sample) which could explain that we find similar results. Notice also that for galaxies with $log(M_*/M_\odot)>11.2$, the contribution from disky galaxies is almost zero and therefore selecting ETGs or ellipticals has almost no effect in the size evolution. The strong size evolution found for these galaxies is therefore robust to galaxy selections and will be discussed in detail in section~\ref{sec:disc}. 
 

 \begin{figure*}
\includegraphics{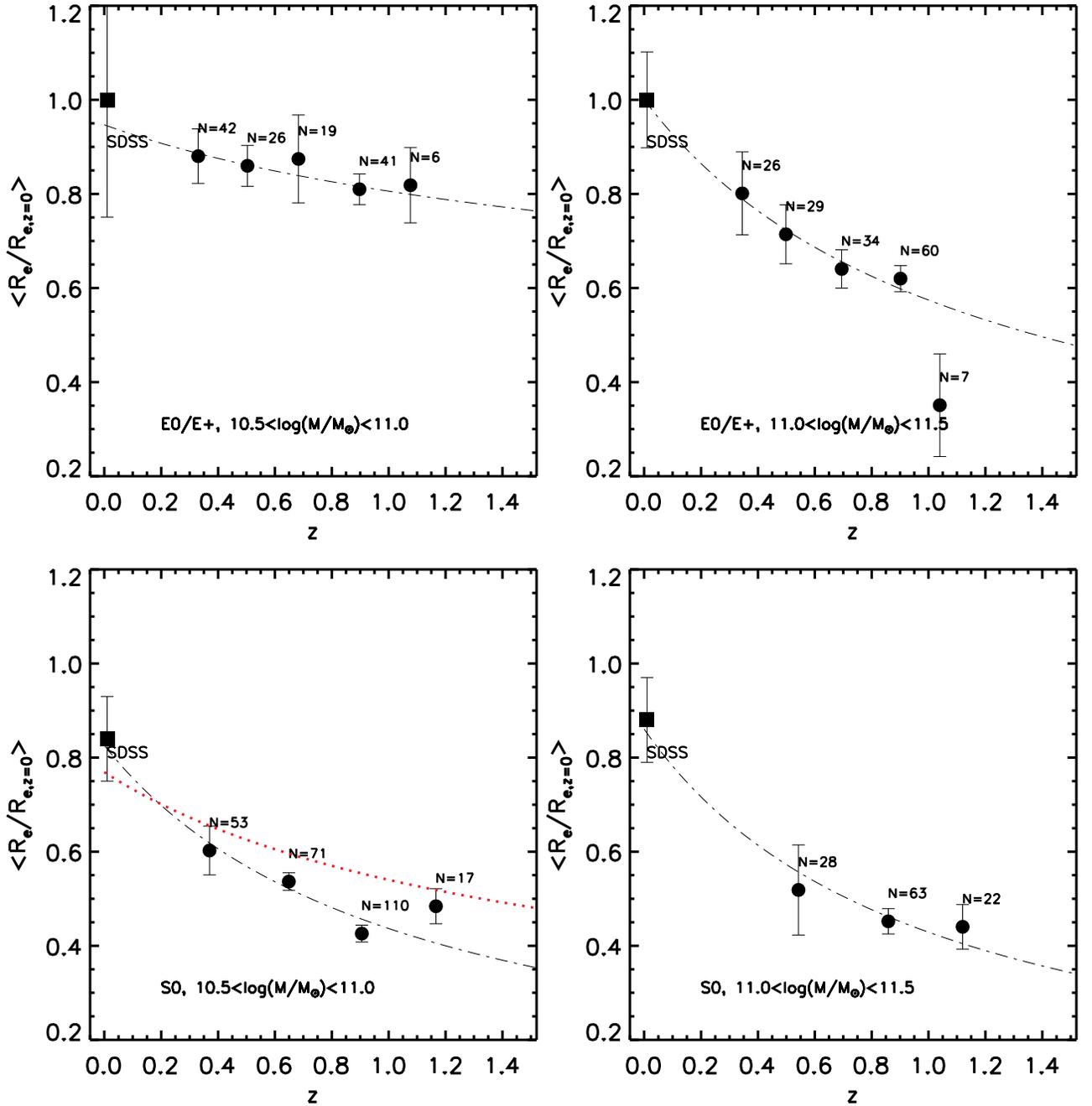}
\caption{Size evolution of passive elliptical (top) and S0 (bottom) galaxies in two mass bins. Left panel: $10.5<log(M/M_\odot)<11$, right panel: $log(M/M_\odot)>11.2$. Dotted-dashed lines are the best fit lines and red dotted line in the bottom left panel is the measured evolution of star-forming galaxies (see text for details). Numbers indicate the number of objects in each redshift bin. Error bars are the scatter of the distributions. SDSS points are computed using the catalogs from Nair \& Abraham (2010) and Simard et al. (2011). }
\label{fig:evol_obser}
\end{figure*}
 
   \begin{figure*}
\includegraphics[scale=.75]{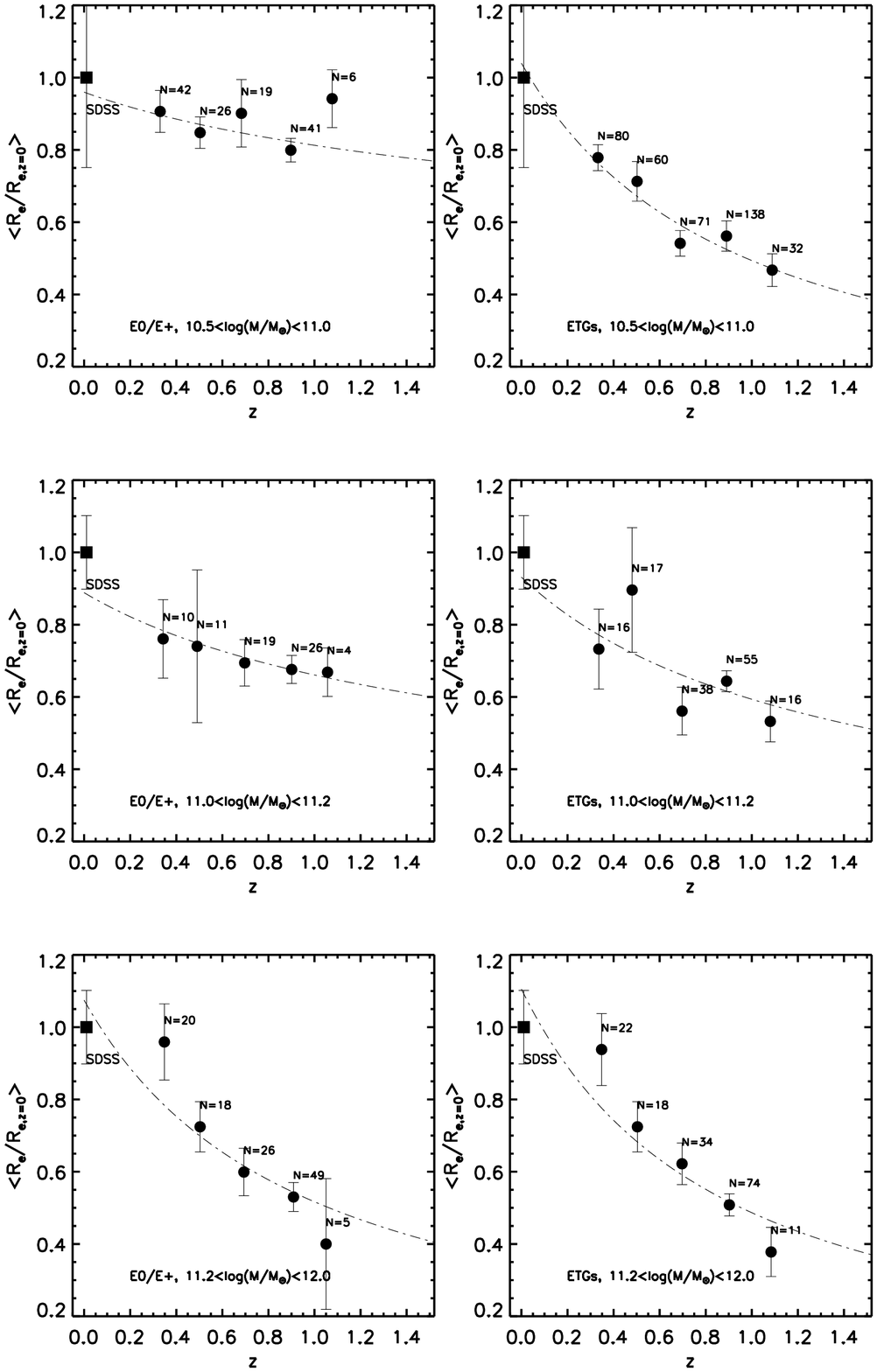}
\caption{Size evolution of elliptical (left column) and early-type passive galaxies (right column) in three mass bins. Top row: $10.5<log(M/M_\odot)<11$, middle row: $11.0<log(M/M_\odot)>11.2$ and bottom row: $11.2<log(M/M_\odot)>12$. Numbers indicate the number of objects in each redshift bin. Error bars are the scatter of the distributions.}
\label{fig:evol_ETG}
\end{figure*}
 
\subsubsection{Lenticulars and early spirals}
\label{sec:s0s}

 The bottom panels of Fig.~\ref{fig:evol_obser} show the size evolution of lenticular galaxies in two mass bins normalized to the same local relation. They appear on average more compact than ellipticals, in a similar way as in the local Universe, as shown in section~\ref{sec:morpho} and confirming the fact already pointed out by \cite{2011ApJ...730...38V} that the most compact galaxies at high redshift have a disk component (this effect is preserved even when not circularizing the radii). Since we normalize with the same local relation as for ellipticals, we also plot, for reference, the ratio of lenticular sizes over those of the ellipticals in the SDSS (notice that all fractions are normalized to the local SDSS elliptical sizes).

 What is interesting is that the size growth of lenticulars does not seem to depend significantly on stellar mass given the large uncertainties, contrary to the behavior shown for the elliptical population: $r_e\propto(1+z)^{-0.67\pm0.30}$ and $r_e\propto(1+z)^{-1.02\pm0.25}$ for low mass and massive lenticulars, respectively.  
 

In order to further investigate what is driving the size increase in these galaxies, we made a simple exercise which is to compare the observed evolution to the one expected in a star-forming disk dominated population. We selected for that purpose, galaxies in our sample with $M(NUV)-M(R)<3.5$ and a spiral-like morphology ($max(P(E), P(S0), P(Sab), P(Scd))=P(Sab)$ or $max(P(E), P(S0), P(Sab), P(Scd))=P(Scd)$), computed the size growth as a function of redshift as we did for the S0s. Since the overall sizes of late--type galaxies are larger than ETGs (see also \citealp{2012arXiv1205.1785M}),  we divide the overall relation by a factor of $1.3$ so that the values at $z=0$ agree (see fig.~\ref{fig:evol_obser}). We just plot the relation for the low mass end because there are not enough high mass late-type to derive a reliable relation at high masses.  

Even though the normalization for disks is still slightly higher, the trends of the size growth for the two populations are very similar ($r_e\propto(1+z)^{-0.67\pm0.30}$ for lenticulars and $r_e\propto(1+z)^{-0.54\pm0.44}$ for star forming disks) which suggests that the size growth of a large part of the early--type galaxies and that of the late--type galaxies  evolve at similar rates. If we add to the passive S0 population, also passive early-spirals ($max(P(E), P(S0), P(Sab), P(Scd))=P(Sab)$ and $M(NUV)-M(R)>3.5$) the trend is preserved.

\subsection{Environment}
\label{sec:environment}

We now study the effect of environment on the mass--size relation of galaxies paying special attention to the uncertainties due to the galaxy sample selection. 

Figure~\ref{fig:group_field_ETGs} shows the evolution of mass-normalized radii $\gamma$ for $log(M_*/M_\odot>10.7)$ passive ETGs in groups and in the field. In the right panel we show only group members with a probability greater than 0.5 and within $0.5\times R_{200C}$ from the group center (central selection) to make sure that the signal is not washed out by interlopers. No significant differences are observed between the central and the larger sample.  Our main result is therefore that the mass--size relation does not depend on environment for field and groups with halo masses $M_{200C}/M_{\odot}<10^{14.22}$. Fig.~\ref{fig:re_evol_envir_morpho} also shows the evolution of the size of field and group passive galaxies divided into the same morphological and stellar mass bins than in the previous section. Relations are more scattered because of the Poisson noise (especially for massive lenticulars) but we do recognize the same trends as for the mixed population. 

A point to take into account when normalizing with the local relation is the fact that groups and field galaxies in the local universe could follow different mass-size relations. We have checked this point by comparing the median mass-size relation of early-type galaxies in groups from the \cite{2007ApJ...671..153Y} catalog to the full DR7 sample. \cite{2007ApJ...671..153Y} put together a catalog of $\sim300000$ groups detected in the SDSS DR4 using an automated halo-based group finder. For this work, we restricted to groups with more than 2 members and removed those objects affected by edge effects ($f_{edge}<0.6$). According to their figure~2, $\sim80\%$ of the groups have $\sim20\%$ contamination which is comparable to the expected contamination in our sample. We use as halo mass estimate, HM1, which is based on the characteristic luminosity of the group but results remain unchanged when using an halo mass estimate based on the characteristic stellar mass. The group catalog has been correlated with the catalog of 2D sersic decompositions by Simard et al. (2011) to get a size estimate for all our galaxies (circularized effective radius of the best fit) as well as with the morphological catalog by \cite{2011A&A...525A.157H} to select ETGs ($P_{early}>0.8$). The mass-size relations that we obtained for the field and group galaxies are fully consistent within $1\sigma$,  we therefore use the same local relation to normalize group and field sizes. \\

In \cite{2012ApJ...745..130R}, in a sample dominated by galaxies with $10<log(M/M_\odot)<11$, we have noticed that while differences in median/average sizes cannot distinguish early--type populations in different environments, a test on both the mean and scatter of their distribution can point out environmental differences. 


We extend Kuiper and Kolmogorov Smirnoff tests analysis to our group and field passive early-type galaxy size distribution in different mass bins to test environmental dependences on their distribution scatter that might distinguish size evolution in the groups and the field. We divide our sample in 3 redshift bins ($0.2<z<0.5$, $0.5<z<0.8$ and $z>0.8$) and 2 mass bins ($10.5<log(M/M_\odot)<11$ and $log(M/M_\odot)>11$) and performed the tests for the two group selections.



\begin{figure*}
\includegraphics[scale=0.7]{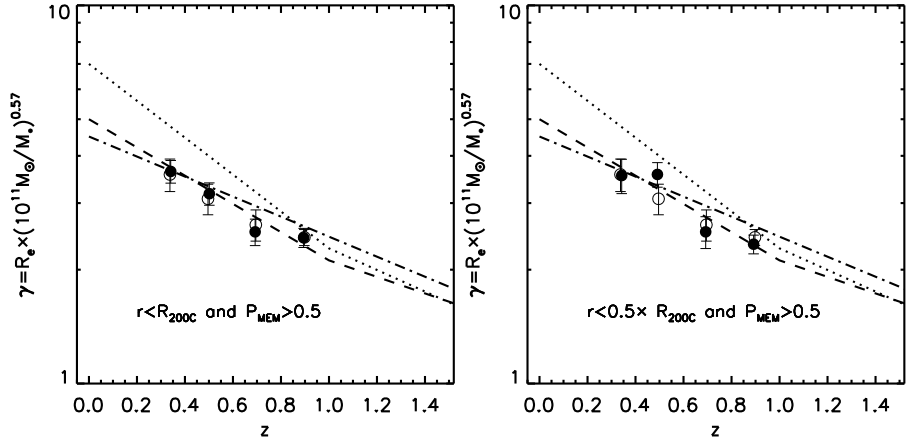}
\caption{Mass-normalized radius for quiescent ETGs in different environments for galaxies with masses $log(M/M_\odot)>10.7$,  as a function of redshift. Filled circles are ETGs living in groups and empty circles are in the field. The left panel shows all group members within $r<R_{200C}$ and a probability to be in the group larger than 0.5 and the right shows the result with a more conservative group selection, i.e. $r<0.5\times R_{200C}$ and $P_{MEM}>0.5$. Dashed line shows Cimatti et al. (2012) fit, dotted-dashed line is Newman et al. (2012) and dotted line is Damjanov et al. (2011). Samples selected using the S\'ersic index tend to show larger sizes, because of the contamination from passive spirals, which is larger in field samples. However, these differences are within 1$\sigma$.}
\label{fig:group_field_ETGs}
\end{figure*}

Results of the analysis are summarized in table~\ref{tbl:stats} and the different size distributions for several sample selections are shown in Fig.~\ref{fig:size_ratios}. Same results for the central selection are shown in the appendix. For all redshift and mass bins, both the KS and Kuiper test results almost always give a probability $> 80\%$ that field and groups size ratios are drawn from the same distribution even when considering only group members with $P_{MEM}>0.5$ and close to the group center.



 Recent works at similar redshifts have found that galaxy size evolution depends on environment. \cite{2011ApJS..193...14C} find indeed that denser environments at $z<1$ tend to be populated by larger galaxies. Differences with this work can come from the way environment is measured and/or how galaxy selection is performed. Cooper and collaborators selection is indeed based exclusively on the S\'ersic index, which as shown in Fig.~\ref{fig:mass_size_sersic} leads to the inclusion of some star-forming galaxies which tend to increase the scatter of the field population. As a matter of fact, we find that KS and Kuiper tests applied to the size distributions of group and field galaxies selected on the basis of the S\'ersic index do show that the distributions have probabilities $<50\%$ to be drawn from the same distributions. However, median sizes do not change significantly so this fact cannot fully explain the difference between the two works. Concerning the environment measurement, we use here the DM halo mass as primary environment estimator whereas Cooper and collaborators use the local density based on neighbors.


\begin{figure*}
\includegraphics[scale=1]{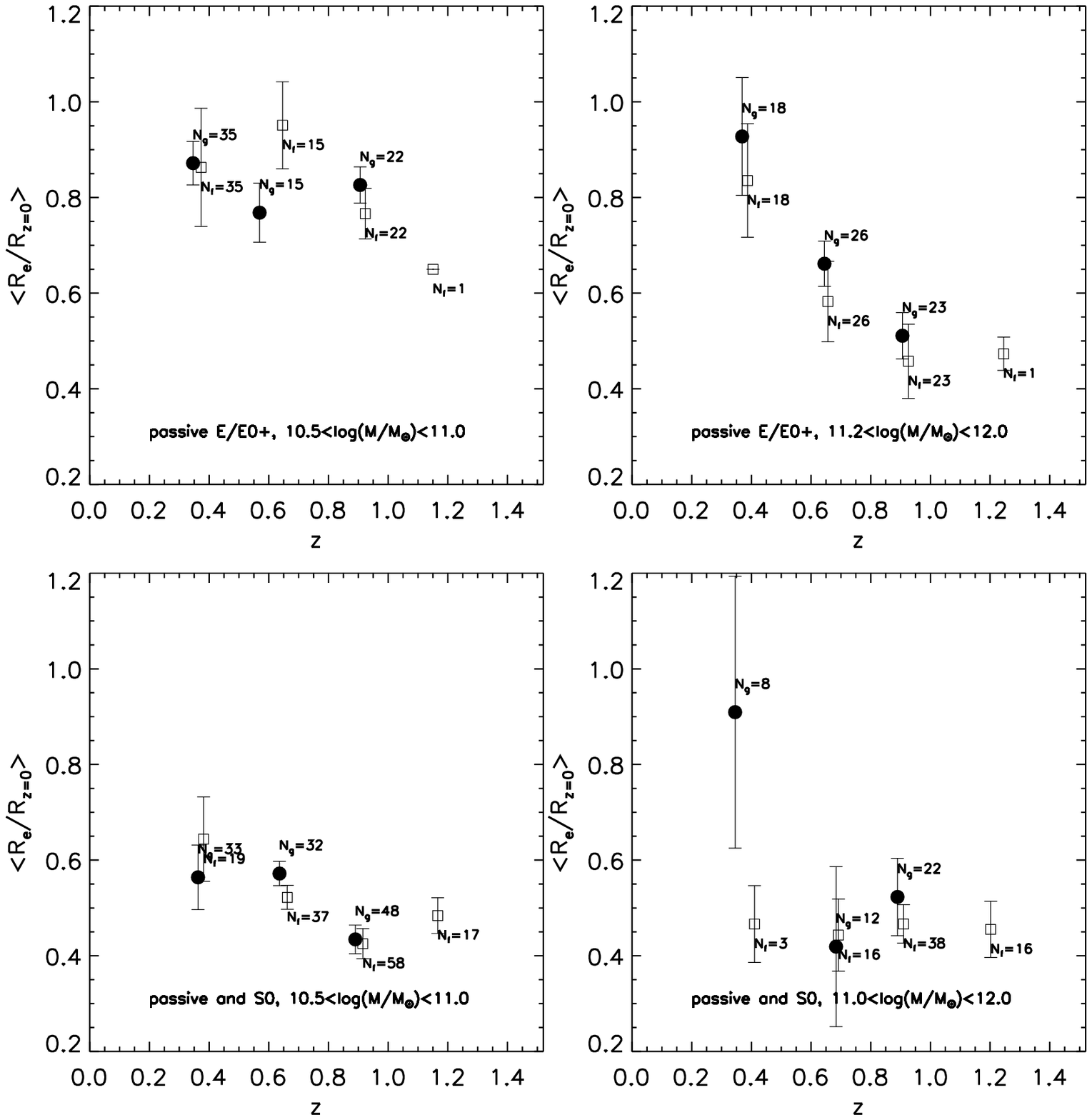}
\caption{Evolution of the size of galaxies from our sample as a function of redshift referred to the local relation from SDSS (see text for details). Filled circles are group galaxies and empty squares are from the field. Error bars are related to the scatter of the size distribution (see text) and numbers indicate the number of galaxies in each redshift bin in the group sample. }
\label{fig:re_evol_envir_morpho}
\end{figure*}

\begin{figure*}
\includegraphics[scale=0.7]{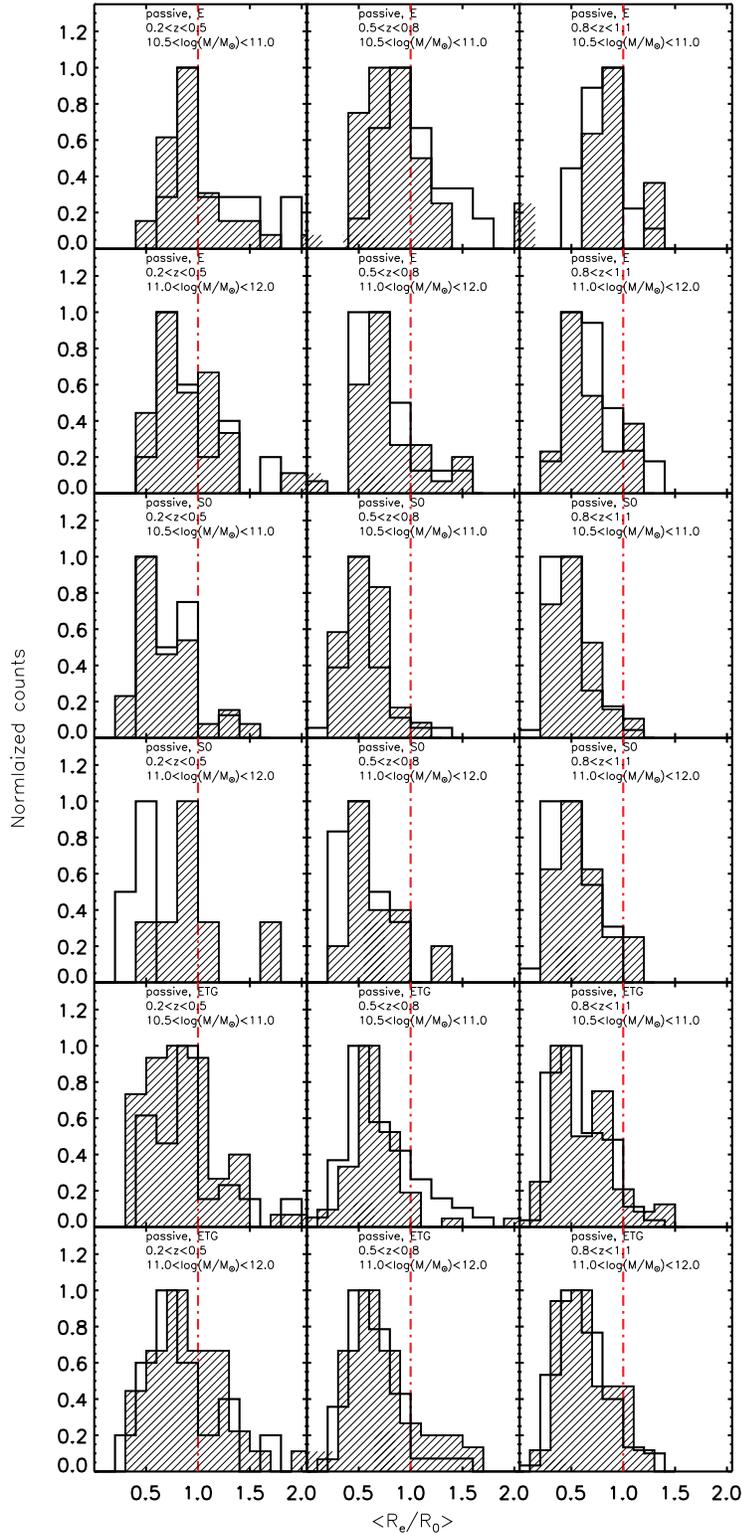}
\caption{Size ratio distributions of group (filled histograms) and field galaxies (empty histograms) for different sample selections and different redshift and stellar mass ranges. The red dashed vertical line indicates a size ratio of 1.  Each column shows a different redshift bin ($0.2<z<0.5$, $0.5<z<0.8$ and $0.8<z<1.1$ from left to right). Each row shows a different morphology and mass selection as explained in the text. }
\label{fig:size_ratios}
\end{figure*}

\begin{table*}
\begin{tabular}{ccccccc}
\hline\hline\noalign{\smallskip}
Redshift & $log(M/M_\odot)$ & Sample & K-S & Kuiper & $N_{group}$ & $N_{field}$\\ 
\hline\hline\noalign{\smallskip}
$0.2<z<0.5$ & 10.5-11 & Pall--1 & 0.95 & 0.99 & 68 & 36\\
& & Pall--2 & 0.85 & 0.97 & 33 & 36\\
 &  & PEs--1 & 0.98 & 0.98 & 35 & 17\\
  &  & PEs--2 & 0.34 & 0.51 & 17 & 17\\
  &  & PS0s--1 & 0.75 & 0.96 & 33 & 19\\
  &  & PS0s--2 & -- & -- & 6 & 19\\
  & 11-12 & Pall--1 & 0.99 & 0.97 & 38 & 16 \\
  & & Pall--2 & 0.99 & 0.99 & 22 & 16\\
  &  & PEs--1 & 0.99 & 0.99 & 30 & 13\\
  &  & PEs--2 & 0.99 & 0.99 & 17 & 13\\
    &  & PS0s--1 & -- & -- & 8 & 3\\
    &  & PS0s--2 & -- & -- & 1 & 3\\
  $0.5<z<0.8$ & 10.5-11 & Pall--1 & 0.43 & 0.71 & 47 & 57 \\
  & & Pall--2 & 0.12 & 0.85 & 19 & 57\\
   &  & PEs--1 & 0.92 & 0.99 & 15 & 20 \\
   &  & PEs--2 & 0.40 & 0.78 & 7 & 20 \\
      &  & PS0s--1 & 0.99 & 0.98 & 32 & 37\\
      &  & PS0s--2 & -- & -- & 6 & 37\\
       & 11-12 & Pall--1 & 0.80 & 0.98 & 49 & 39\\
       & & Pall--2 & 0.87 & 0.99 & 26 & 39\\
        &  & PEs--1 & 0.93 & 0.99 & 37 & 23 \\
        &  & PEs--2 & 0.92 & 0.95 & 20 & 23 \\
          &  & PS0s--1 & 0.12 & 0.15 & 12 & 16\\
           &  & PS0s--2 & -- & -- & 1 & 16\\
          
           $0.8<z<1.0$ & 10.5-11 & Pall--1 & 0.99 & 0.99 & 70 & 82\\
           & & Pall--2 & 0.42 & 0.99 & 24 & 82\\
   &  & PEs--1 & 0.88 & 0.99 & 22 & 24 \\
      &  & PEs--2 & 0.88 & 0.99 & 9 & 24 \\
      &  & PS0s--1 & 0.99 & 0.99 & 48 & 58\\
        &  & PS0s--2 & 0.88 & 1.00 & 10 & 58\\
       & 11-12 & Pall--1 & 0.95 & 0.70 & 53 & 89\\
       & & Pall--2 & 0.55 & 0.98 & 29 & 89\\
        &  & PEs--1 & 0.95 & 0.99 & 31 & 51\\
        &  & PEs--2 & 0.54 & 0.98 & 21 & 51 \\
          &  & PS0s--1 & 0.55 & 0.91 & 22 & 38\\
           &  & PS0s--2 & -- & -- & 7 & 38\\

 \hline

\end{tabular}
\caption{Results of Kuiper and Kolmogorov-Smirnoff statistical tests applied to group and field galaxies size ratio distributions in different samples (see text for details). Pall stands for all passive galaxies, PEs are passive ellipticals and PS0s passive lenticulars. $N_{group}$ and $N_{field}$ indicate the number of galaxies in the group and field samples respectively. Numbers ($1$ and $2$) indicate if the group sample is taken at $r<R_{200C}$ ($1$) or at $r<0.5\times R_{200C}$ (see text for details). We did not compute the statistical tests when the number of objects was below 10.}
\label{tbl:stats}
\end{table*}


\subsubsection{BGGs}

Brightest group galaxies (BGGs) deserve a particular mention since they have been more studied because of their very particular position at the center of massive structures and their high surface brightness which make them easy to detect and analyze. Current works, though, do not agree about the BCG size evolution and how BCG sizes compare to those of field and satellite galaxies. 

Based on SDSS data, \cite{2009MNRAS.395.1491B} found that BCGs are larger than field and satellite galaxies at fixed stellar mass and that there is a steep evolution of their size from $z\sim0.3$ to present. The author also argues that minor dry mergers are the most probable mechanism to explain the build-up of these objects. Also in the SDSS, \cite{2009MNRAS.394.1213W} did not find a significant difference between the sizes of centrals and satellites in groups. 

At higher z, \cite{2011ApJ...726...69A} find a significant size evolution from $z\sim0.6$ to present, but do not detect any evolution in their profile. They interpret this fact as a signature of feedback instead of merging. \citep{2011MNRAS.414..445S} studied a sample of high redshift BCGs and found a very mild evolution of the BCG size from $z\sim1$. 

From the modeling point of view, \citet{2011arXiv1105.6043S} showed that BCGs should evolve much faster than satellite galaxies. 

We use the BGGs definition as the most massive galaxies within an NFW scale radius of the X-ray position, from the George et al. (2011) catalog. Fig.~\ref{fig:bcgs} shows the size evolution for passive BGGs and satellite group members with similar stellar mass ($log(M/M_\odot>11$). We find that the two population evolve in a similar way within the error bars, e.g. we do not observe significant differences in the size evolution of BCGs as compared to satellite group members with similar stellar mass.

\begin{figure*}
\includegraphics[scale=1]{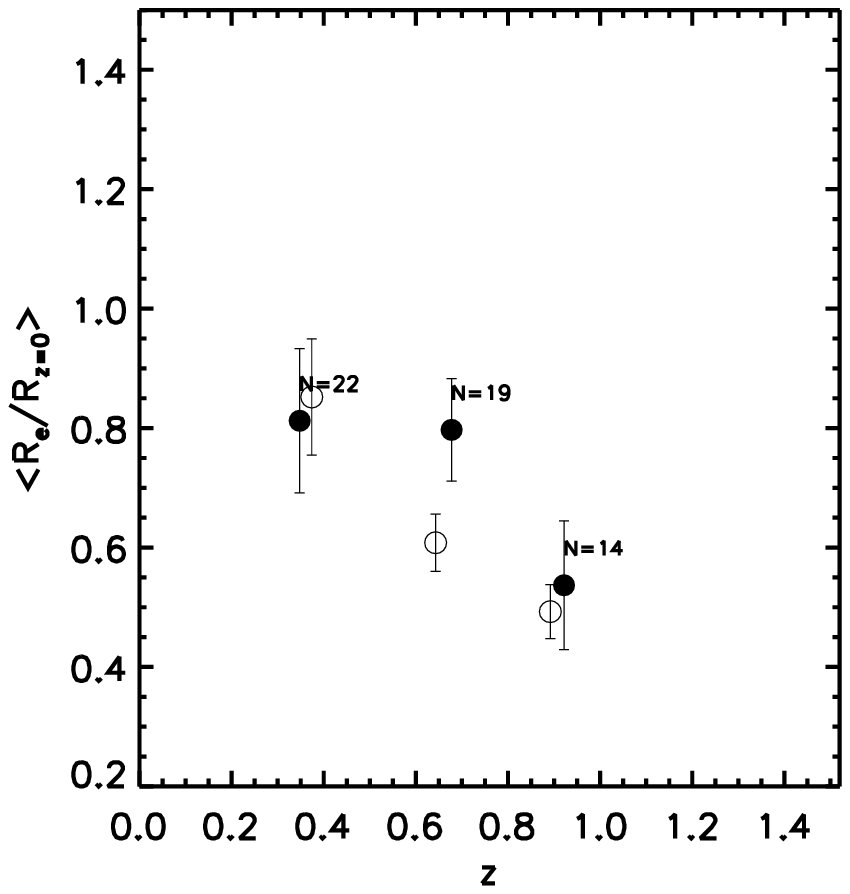}
\caption{Size evolution of BCGs (filled circles) compared to satellite group members with $log(M/M_\odot)>11$ (empty circles). Numbers show the number of BCGs in each bin.}
\label{fig:bcgs}
\end{figure*}

\section{Discussion}
\label{sec:disc}
To understand why not all populations evolve in the same way, we consider different scenarios of galaxy growth. There are two main, well distinct, physical mechanisms proposed so far in the literature
to puff up massive bulges from high redshifts
to the local Universe: galaxy mergers and mass loss
via quasar and/or stellar wind. 

\subsection{Mergers}

Hierarchical models of structure formation 
envisage the growth in size of massive galaxies 
via a sequence of major and minor
mergers. While (mainly gas-rich) major mergers are believed to happen at high redshifts
and may be responsible for forming the galaxy, minor dry merger happen
on cosmological timescales and tend to impact mainly
the external regions of the galaxy, thus increasing its size, 
but leaving its inner regions mostly unaltered.

In this work we consider the predictions of several, representative,
hierarchical galaxy evolution models that 
differ in terms of underlying techniques and physical assumptions.
We adopt, more precisely, \citet{2006MNRAS.370..645B}, \citet{2009MNRAS.398..898H}, \citet{2010MNRAS.404.1111G} and \citet{2011arXiv1105.6043S}. All model predictions have been computed for the range of stellar masses
\footnote{All predictions have been corrected to a common Chabrier IMF. 
We however expect differences in the IMF
to have a minor impact on the rate of size evolution of massive spheroids.} 
and galaxy type of interest to this paper. We mainly considered galaxies with $B/T>0.5$ 
when comparing to early-type galaxies (though we checked that our
results are practically unchanged when restricting to $B/T>0.7$), and galaxies with $0.3<B/T<0.7$
when comparing to S0s.

Both the \citet{2006MNRAS.370..645B}} and \citet{2010MNRAS.404.1111G} models follow the hierarchical
growth of galaxies along the merger trees of the Millennium simulation \citep{2005Natur.435..629S}.
Galaxy progenitors are initially disk-like and after a major merger the
remnant is considered to be an elliptical (though disk regrowth can happen). 
Minor mergers instead tend to preserve the initial morphology of the most massive
progenitor but tend to increase the mass of the bulge and disk components
via the aggregation of old stars and newly formed ones during the merger.
Half-mass sizes are then updated at each merger event assuming energy conservation.
Despite being built on the same dark matter simulation,
the subhalo/galaxy merger rates of these two models differ due to the different corrections in dynamical
friction timescales. In both models, bulges can also grow via disk instabilities.
However, the implementation of the latter physical ingredient substantially differs in the two
models, with \citet{2006MNRAS.370..645B} assuming a much stronger bulge growth via disk instabilities
with respect to the \citet{2010MNRAS.404.1111G} model (see discussion in \citealt{2012A&A...540A..23S}).
Most importantly for size evolution, both models do not consider gas dissipation during gas-rich major mergers,
a feature that has instead been included by \citet{2011arXiv1105.6043S} by properly adapting the
\citet{2010MNRAS.404.1111G} model.
The \citet{2009MNRAS.398..898H} model follows the analytic mass accretion histories of haloes
and at each time step initializes central galaxies and infalling satellites
according to empirical correlations inspired by halo occupation techniques and
high-redshift data. Equivalently to the models discussed above,
at each merger the half-mass radius is updated following energy conservation
arguments, with also gas dissipation (see also
\citealt{2012MNRAS.tmp.2680N} for a more recent work adopting similar techniques).
\citet{2009MNRAS.398..898H} have mainly focused on early-type galaxies (not lenticulars)
with stellar masses above $M_{*}> 10^{10}{\, \rm M_{\odot}}$.

\subsubsection{The Size Growth of Ellipticals compared to model predictions}

\begin{figure*}
\includegraphics{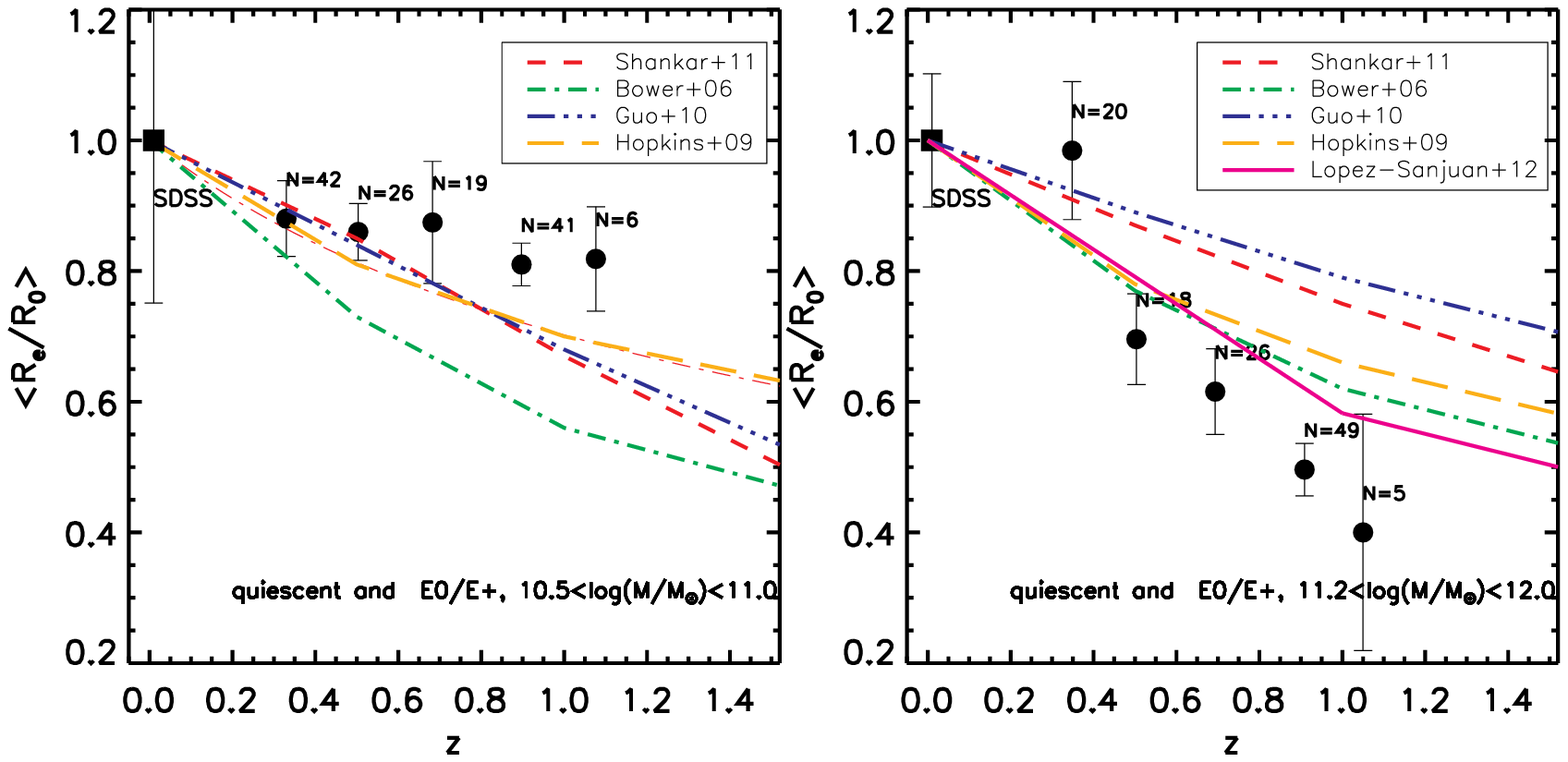}
\caption{Size evolution of elliptical passive galaxies in two mass bins. Left panel: $10.7<log(M/M_\odot)<11$, right panel: $log(M/M_\odot)>11.2$. Numbers indicate the number of objects in each redshift bin. Error bars are the scatter of the distributions. The different lines (see legend) show the prediction from different semi-analytic and empirical models (see text for details). }
\label{fig:ell_evol_models}
\end{figure*}

In Figs.~\ref{fig:ell_evol_models}, we compare our results with predictions from Shankar et al. (2011), Bower et al. (2006) and Guo et al.  (2010). For ellipticals, we focus here on the lowest and highest stellar mass bins of fig~\ref{fig:evol_ETG}, where the differences are stronger (i.e. $10.5<log(M/M_\odot)<11$ and $log(M/M_\odot)>11.2$).
The first relevant feature arising
from the comparison is that most of the theoretical predictions are
in agreement with the (mild) size evolution of ETG galaxies with
stellar masses $M_{*}<10^{11}$ (left panel of Fig.~\ref{fig:ell_evol_models}) (specially taking into account the uncertainties due to morphological classifications). Merger models are therefore successful to predict the size evolution of these lower mass ellipticals. 

More interestingly, despite the significant variance in the input parameters and/or physics most merger models seem to be unable to reproduce completely the
fast drop in sizes for quiescent ellipticals with stellar masses $Log(M_{*}/M_\odot)> 11.2$, especially at $z> 0.5$ (right panel of Fig.~\ref{fig:ell_evol_models}).
This is in line, and further complements, the recent claims of a possible
inefficiency of the puffing-up via mergers pointed out by \cite{2011arXiv1105.6043S}, \cite{2012MNRAS.tmpL.424C}, and \cite{2012MNRAS.tmp.2680N}. Notice that in this stellar mass bin, the contribution of S0 galaxies is almost negligible (see fig.~\ref{fig:evol_ETG}), so a similar behavior is found when selecting all ETGs with $log(M_*/M_\odot)>11.2$ and should be independent of the morphological selection.

\begin{figure*}
\includegraphics[scale=1]{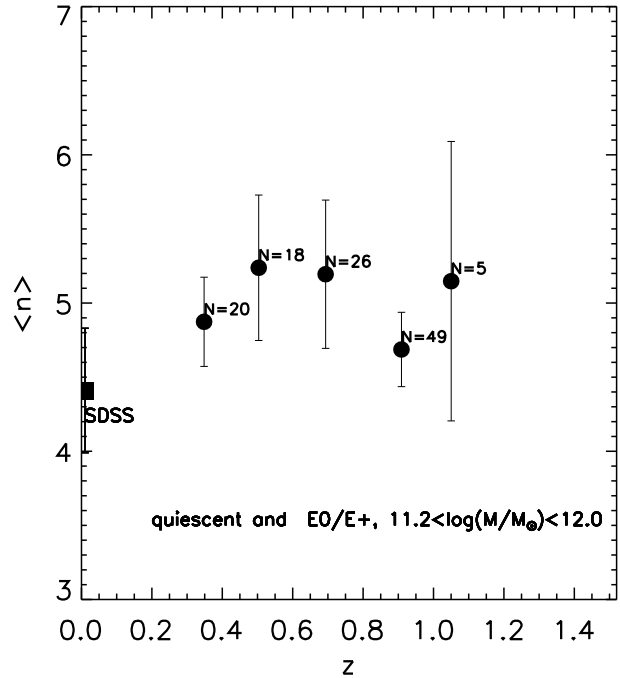}
\caption{Median S\'ersic index evolution for elliptical galaxies in the mass range $11.2<log(M/M_\odot)<12$. The local point comes from the S\'ersic decompositions of Simard et al. (2011) on visually classified ellipticals by Nair \& Abraham (2010). }
\label{fig:sersic_evol}
\end{figure*}

\citet{2012MNRAS.tmp.2680N} have recently shown the results of a merger model initialized
via halo occupation techniques that,
by neglecting dissipative processes and assuming only mergers
with spheroids, maximizes the evolution in surface density. 
They conclude that minor and major mergers 
may not be sufficient to explain the observed size growth of early type galaxies (see also \citealp{2012arXiv1202.4674L}).  An other work where the size growth since $z\sim1$ is explained almost completely by mergers \citep{2012ApJ...746..162N} has to assume very short merger time scales ($1$ Gyr) and steep growth efficiency, optimistic with respect to observations (notice  in addition that the size evolution they measure is less steep than the one reported here though, because of the selection used, as shown in sections~\ref{sec:sel} and \ref{sec:growth}). \citet{2012ApJ...744...63O} have recently analyzed 40 cosmological re-simulations of individual massive
galaxies with final $M_{*}>6.3 \times 10^{10} {M_{\odot}}$, out of which 25 appear quiescent early-type galaxies.
While they claim a strong size evolution in the \emph{cumulative} distribution
of galaxies with present stellar mass $M_{*}> 6.3 \times 10^{10} {M_{\odot}}$,
a close inspection of their Figure 1 (left panel) reveals that galaxies above
$M_{*}> 2 \times 10^{11} {M_{\odot}}$ have indeed had a rather mild 
size evolution at fixed stellar mass of about $\sim 30-40 \%$ at $z< 1$. Nevertheless, this may not necessarily reflect a failure of hierarchical models, as part of
the discrepancy could simply arise from the specific underlying assumptions made. For example, the early semi-analytic model by \citet{2006ApJ...648L..21K} 
predicts a stronger size evolution for the very
massive galaxies with $M_{*}> 5\times 10^{11}{\times M_{\odot}}$, 
in better agreement with observations. 

Overall, most merger models have some difficulties in fully reproducing the size evolution of the most massive early-type galaxies.
Part of the apparent evolution in sizes may be driven
by relatively younger larger galaxies formed at later epochs (progenitor bias) even if the relevance of this effect is still unclear. A recent empirical model by \cite{2012arXiv1202.4674L}, based on merger observations in the COSMOS field, shows in fact that taking into consideration major and minor mergers observed in COSMOS can explain $\sim 55\%$ of the size evolution of massive ($>10^{11} {M_{\odot}}$) ETGs. If the progenitor bias of massive ETGs accounts for a factor 1.25, this work can explain $\sim 75\%$ of the size evolution.  We also show, for completeness, in figure~\ref{fig:ell_evol_models} the expected evolution taking into account this effect from that work (notice however that the selection in terms of morphology and star-formation is not exactly the same than the one used in this work). However, \cite{2012ApJ...745..179W} showed that the recently quenched galaxies at $1<z<2$ are not significantly larger than their older counterparts, suggesting that the effect of the so called progenitor bias is limited. Moreover, fig.~\ref{fig:morpho_fractions} also shows that the number densities (fractions) of massive ellipticals do not evolve significantly from $z\sim1$, pointing again towards a reduced effect of newly formed galaxies.

%


There are also other possible tensions
between merger models and current observations. A puffing-up of the external regions
with no inner density variation as suggested by mergers \citep[e.g.,][]{2009ApJ...699L.178N},
naturally produces an increase in the S\'{e}rsic index \emph{n} of an ideally light profile fitted to
the projected density profile. In Fig.~\ref{fig:sersic_evol}, we plot the median S\'ersic index for massive elliptical galaxies. 
The median S\'ersic index of the overall population does not evolve since $z\sim1$. These results 
are in possible tension with this prediction
(similar conclusions were found by \cite{2011MNRAS.414..445S} for instance).

Another prediction of merger models is that galaxies residing at the centre
of more massive haloes should, at fixed stellar mass, experience more mergers
and thus be larger than their counterparts in less massive haloes
, at least above $M_h>5 \times 10^{12} M_\odot$, according
to the analysis of \citet{2011arXiv1105.6043S}. The latter, in fact, showed that BCGs should evolve
 much faster
at $z<2$, and also end up being larger
than other galaxies of similar mass. We do not observe this trend.

\subsubsection{Environmental dependencies compared to model predictions}

\begin{figure*}
\includegraphics[scale=.7]{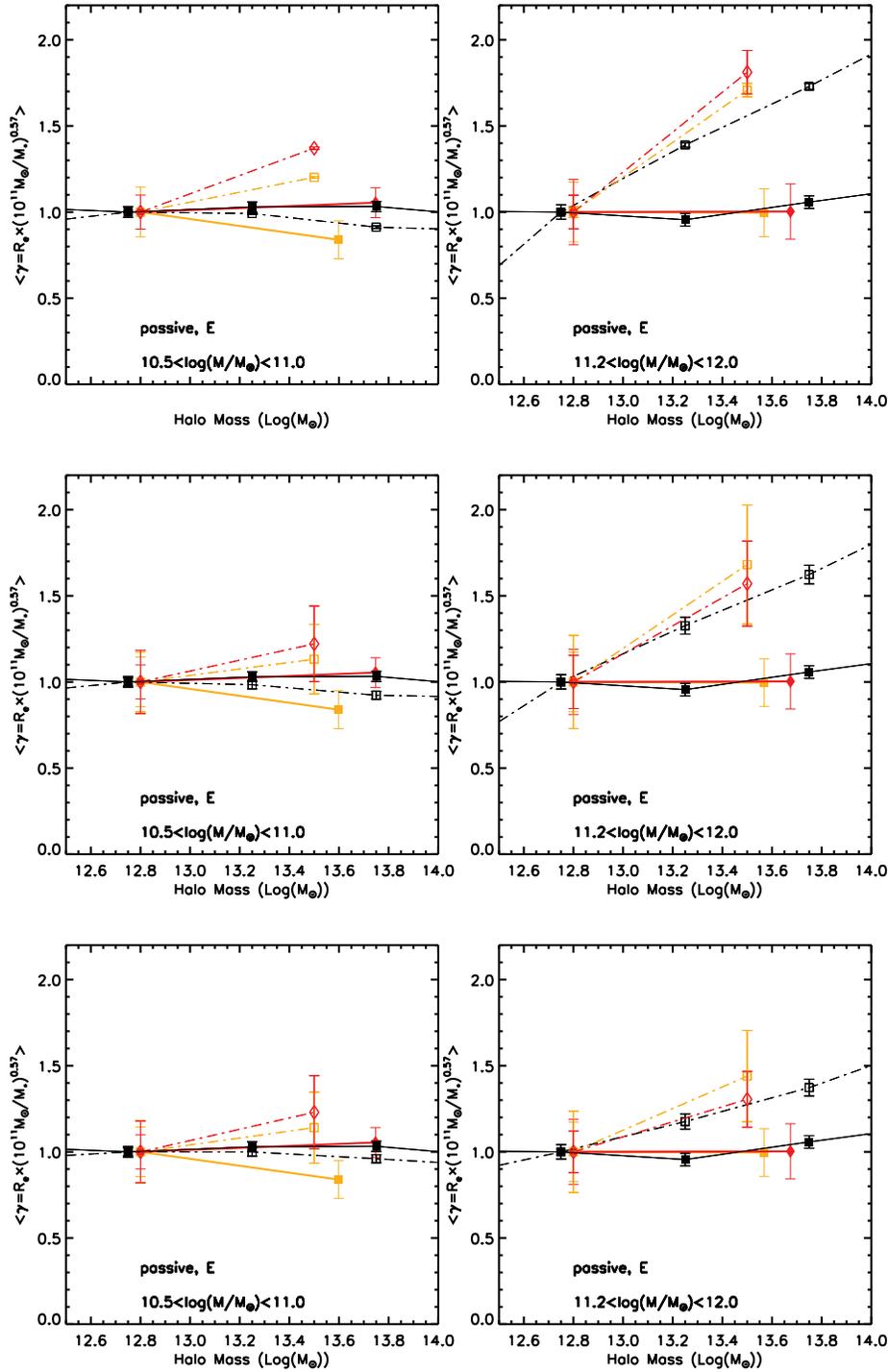}
\caption{Size of elliptical galaxies as a function of halo mass for different redshift bins. Red diamonds are objects with $0.8<z<1.0$, orange squares are $0.5<z<0.8$. The black dashed-dotted line, the orange dashed line and the red dashed-three dotted line show the expected relation from Shankar et al. (2011) models at $z=0$, $z=0.5$ and $z=1$. Values from models have been normalized so that, by definition, the field observed value at an halo mass of $log(M_h/M_\odot)=12.8-13$ is equal to 1. The solid black line and the filled squares are the values measured from the SDSS and the group catalog from Yang et al. (2007). First row shows model prediction without errors, the second row shows results when the number of objects in models are selected to match the observations and $\sim30\%$ field contaminations are included, finally the third row also includes a $30\%$ error in halo mass. Errors in models and observations are errors on the median values computed through bootstrapping (see text for details).}
\label{fig:halo_size}
\end{figure*}

 Concerning the dependence of galaxy sizes and their evolution on environment, 
we compare our observations to predictions from Shankar et al. (2012) in figure~\ref{fig:halo_size}.  We plot the mass--normalized radius as a function of halo mass for two redshift bins from COSMOS ($0.5<z<0.8$ and $0.8<z<1.0$), as well as for the local Universe from SDSS, for two mass ranges. All radii  are shown in units of the mass--normalized radius measured in the field.
Uncertainties in the observations have been calculated by bootstrapping 1000 times and recomputing the median of the distribution each time, as in all previously shown plots. 

As discussed in Section~3 and summarized in Figs.~\ref{fig:re_evol_envir_morpho},~\ref{fig:group_field_ETGs} and~\ref{fig:halo_size}, our COSMOS analysis does not show any significant dependence of median galaxy size on large-scale environment, e.g., galaxy sizes have similar medians in the field and in the groups. We observe the same in the SDSS sample, comparing measurements in galaxy groups from Yang et al. (2007) with the field. 

The top panels of figure~\ref{fig:halo_size} compares observations to model predictions with no accounting for our sample size and observational uncertainties. In this case, the Shankar et al. (2012) model predicts that our most massive halos should have mass-normalized radii of close to twice that of the field.

The middle and lower panels demonstrate that this strong difference is diluted when taking into account uncertainties intrinsic to our sample: the number of galaxies in each bin, the uncertainty due to photometric redshift estimation, and that on the estimate of the halo mass. In the middle panel, to properly compare observations with model predictions at each redshift and halo mass interval of interest, we perform
1000 Monte Carlo realizations in which we draw subsamples of galaxies from the \cite{2011arXiv1105.6043S} catalog with numbers equal to those in the SDSS and COSMOS samples.
Galaxies are selected to have $B/T>0.5$ and to share the
same stellar and halo mass intervals as in the observations, and sizes have been normalized to the local mass-size relation \cite{2011arXiv1105.6043S}.
To each mock subsample we substitute 30\% (when comparing to COSMOS, \citealp{2011arXiv1109.6040G}) and
20\% (when comparing to SDSS, \citealp{2007ApJ...671..153Y}) of members with galaxies of the same stellar mass residing 
in the halo bin with the lowest mass in order to mimic contamination from the field (e.g. because of photometric redshift uncertainties).  As expected, this tends to reduce the increase of mean size with halo mass by up to 20\%.  For each Monte Carlo realization, we compute the mean and then extract, from the full distribution of means, the final mean and its 1-sigma error. In the lower panel, we add the further uncertainty due to halo mass estimation. We  included in the simulations a Gaussian scatter of $0.3$ dex width to reproduce the average uncertainties in the halo mass (Yang et al.  2007; Leauthaud et al. 2010).

When taking into account all these sources of uncertainties (lower panels of figure~\ref{fig:halo_size}), the model predicts a much smaller difference in size between the most massive halos and the field. Even if the trend observed in the upper panels still holds, when comparing our results obtained with COSMOS to the model, they are consistent at $1\sigma$, i.e., the mass-normalized radius does not depend on environment.
However, the larger SDSS sample shows that the model predicts sizes in groups to be about 1.5 times larger than those in the
field, at variance with the observations at more that 3~$\sigma$.  In future work, we will investigate  if this result is specific to the particular model we consider (Shankar et al., in preparation) and/or to the observational data we have used (Huertas--Company et al., in preparation). 

\subsubsection{The Size Growth of Lenticulars compared to model predictions}

\begin{figure*}
\includegraphics{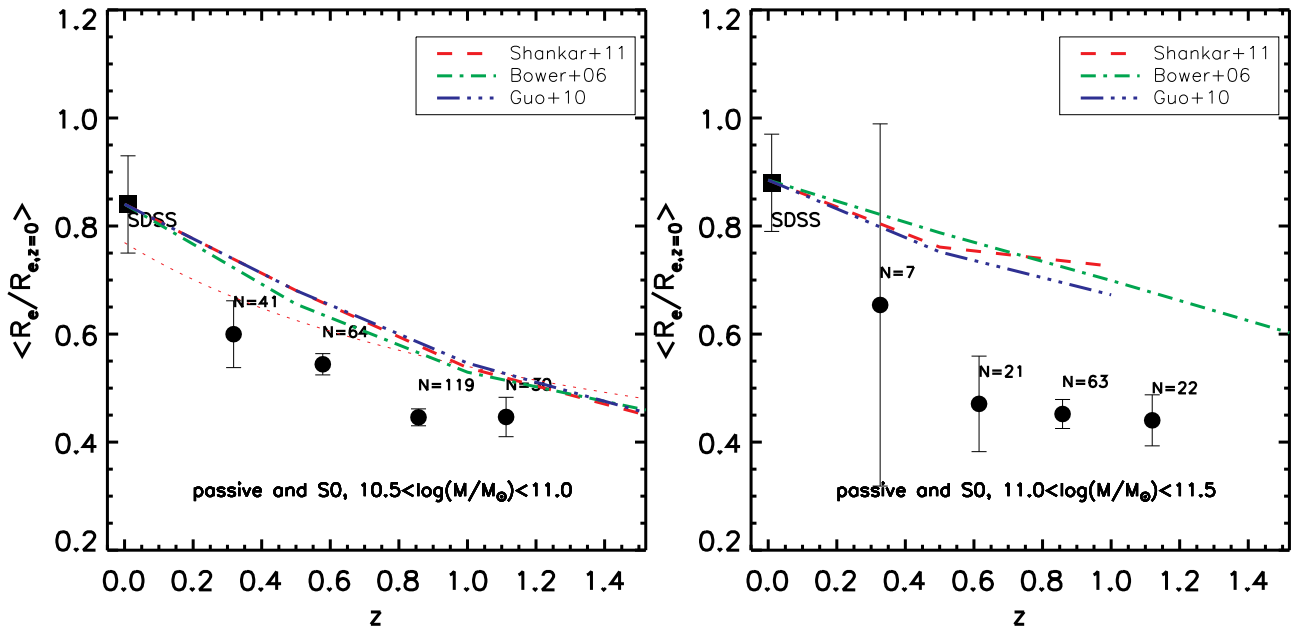}
\caption{Size evolution of lenticular passive galaxies in two mass bins. Left panel: $10.5<log(M/M_\odot)<11$, right panel: $log(M/M_\odot)>11.0$. Numbers indicate the number of objects in each redshift bin. Error bars are the scatter of the distributions. The red dotted line in the left panel indicates the median evolution of star-forming disks in our sample divided by a factor of $1.3$ to reach the same normalization. The other lines (see legend) show the prediction from different semi-analytic models (see text for details). }
\label{fig:sos_evol}
\end{figure*}

For what concerns lenticular galaxies (Fig.~\ref{fig:sos_evol}), model galaxies have been selected to have $0.4<B/T<0.7$.
For this range of $B/T$, disk instabilities also play
a non-negligible role in building bulges (see, e.g., \citealt{2011arXiv1105.6043S}, and references therein).
Despite the significantly different procedures
to grow bulges in the models, from mild instabilities \citep{2010MNRAS.404.1111G} to violent ones \citep{2006MNRAS.370..645B}, the result in the median size evolution is similar to the elliptical results.
It is interesting to note that, analogously to what inferred with respect to
more bulge dominated galaxies, hierarchical models are able to reproduce the trend
in size evolution for lenticulars below $M_{*}< 10^{11}{\, \rm M_{\odot}}$, but they tend
to predict a shallower evolution above this mass.

\subsection{Expansion}
An alternative model
for efficiently puffing up sizes of massive, early-type
galaxies, considers the galaxy expansion consequent to significant  mass loss
via quasar and/or stellar winds \citep[][]{2008ApJ...689L.101F,2010ApJ...718.1460F,2009ApJ...695..101D}.
This model envisages that all massive ($M_{*} > 3 \times 10^{10}{\, \rm M_{\odot}}$) early-type galaxies forming at $z > 1$ 
went through a rapid expulsion of large amounts of mass,
possibly caused by a powerful quasar wind, that caused the galaxy to expand. 
Numerical experiments in favor of this physical scenario have
been performed by \citet{2011MNRAS.414.3690R}, who showed that even in the presence
of large amounts of dark matter, massive galaxies can significantly
expand their sizes by a factor approximately proportional
to the fraction of mass loss. 
\citet{2008ApJ...689L.101F} predict that most of the size evolution for the massive
spheroidal galaxies should be delayed with respect to the peak of quasar activity by about 0.5-1 Gyr.
The model also predicts a milder and possibly longer size evolution for early-type galaxies with stellar
mass below $M_{*} < 2 \times 10^{10}{\, \rm M_{\odot}}$, for which
the dominant energy input in the interstellar medium comes from
supernovae explosions.

Clearly, the typical evolutionary timescales to fully evolve
a galaxy onto the local size-mass relation in the \cite{2010ApJ...718.1460F}
model are in general quite shorter than the cosmological ones
required by a merger scenario. While a fast evolution
in the size growth of massive ellipticals is supported by our data,
their model predicts strong size evolution
mainly at $z > 1$, while our data show strong size
evolution also at  $z<1$, at least for massive ellipticals.

\cite{2010ApJ...718.1460F} also point out that a fast size evolution necessarily should lead to
the co-existence of large and compact quiescent galaxies at
any redshift $z > 1$, and thus a large dispersion up to a factor
of $< 5-6$ in the size distribution at ﬁxed stellar mass. The
dispersion should then significantly reduce below $z < 1$, as
most of the galaxies should be already evolved and their formation rate, which parallels the one of quasars peaking at $z=2$ \citep{2006ApJ...650...42L}, should progressively drop at late times (see their Fig. 1).
The dispersion in sizes for quiescent, massive early-type galaxies that we measure from COSMOS seems instead to be rather contained, within a factor of two at ﬁxed redshift and stellar mass. We also note, however, that our lower mass,
quiescent ETG that have larger sizes might be, at least in part, be composed by
the population of high–redshift, already evolved early-type
galaxies predicted by \cite{2006ApJ...650...42L}.

The expansion model predicts that compact
galaxies should be relatively young at the time of observation, being close to their formation epoch because the size evolution occurs only $\sim20$ Myr after the expulsion via quasar feedback \citep{2011MNRAS.414.3690R}. This
may be possibly at variance with the rather old ages that usually characterize
massive ellipticals (see also \citealt{2011MNRAS.415.3903T} for similar considerations based on observations).

On the other hand, \citet{2011MNRAS.414.3690R} also found in their numerical tests that the galaxy mass profiles
should not change after the blow-out, if the mass loss is contained to a factor of two or so.
This might explain the nearly constancy with redshift of the S\'{e}rsic index in our sample.
Also, being galaxy expansion an in-situ physical mechanism, it should be largely independent
of environment, as suggested by our data (Figs.~\ref{fig:re_evol_envir_morpho} and~\ref{fig:halo_size}).

In summary, while merger models can explain most of the size growth size $z\sim1$, 
we find evidence that both proposed scenarios (mergers and mass loss
via quasar and/or stellar winds) suffer from some shortcomings
with respect to our results and more advances in modeling are clearly needed to deeply understand how the two different scenario shape the galaxy size evolution.


\section{Summary and conclusions}

We have studied a sample of 3146 group galaxies and 3760 field galaxies from the COSMOS survey, and selected 298 group and 384 field galaxies as passive galaxies (based on their M(NUV)-M(R) dust corrected rest-frame color) with $log(M/M_\odot)>10.5$. We show, for the first time, how the mass-size relation and size growth depend on the detailed morphology of the quiescent population (mainly ellipticals and lenticulars) as well as on large-scale environment defined by the dark matter halo mass. \\
Our main results are:

\begin{enumerate}

\item A detailed morphological dissection of the passive population up to $z\sim1$ reveals that: 

\begin{itemize}
\item About $80\%$ of all passive galaxies have an early-type morphology at all stellar masses and at all redshifts from $z\sim1$. The remaining $20\%$ are essentially early-type spirals.
\item Early--type galaxies are both ellipticals and S0s. At all redshifts up to $z\sim1$, elliptical galaxies dominate the ETG population at masses $log(M/M_\odot)>11-11.2$. At $z<0.5$, galaxies with masses $log(M/M_\odot)<11$ are around half ellipticals and half lenticular, while lenticulars dominate the low--mass fractions at $z>0.5$. An important fraction of group and field low-mass passive galaxies in the redshift range $0.5<z<1$ are lenticular galaxies, e.g. have a disk component (see also Bundy et al. 2010, \citealp{2012arXiv1205.1785M})

\end{itemize}

Therefore, studying the population of passive galaxies as a whole mixes different morphological populations which do not necessarily share the same evolutions.\\

\item When separating the ellipticals from the lenticulars, we show that {\it galaxy size evolution strongly depends on mass range and ETG morphology}:

\begin{itemize}
\item Massive ellipticals ($log(M_*/M_\odot)>11.2$) do experience a very strong size evolution from $z\sim1$ to present ($r_e\propto(1+z)^{-0.98\pm0.18}$). Even though the trend is somehow steeper than the one predicted by most of published semi-analytical models, minor dry mergers remain the most plausible explanation to the expansion. However, the S\'ersic index of this population is not significantly evolving with time as expected from hierarchical models. We cannot exclude from our data that expansion via feedback might also play a role on the size evolution of this population. 

\item Less massive ellipticals ($10.5<log(M_*/M_\odot)<11$) do not evolve much from $z\sim1$ (from $\sim10\%$ to $\sim30\%$ depending on which morphological classification we use). This behavior is well reproduced by most of current semi-analytic models from which the size growth is  mainly due to major and minor mergers. 

\item The evolution observed in the lenticular population does not change significantly with stellar mass and they do show a size growth of $55\%\pm10\%$ since $z=1$.
 
\end{itemize}

\item Finally, we studied environmental effects on the size evolution by dividing our sample into field and group galaxies:

\begin{itemize}

\item We do not detect any significant evidence that the evolution of field and group galaxies ($13<log(M_h/M_\odot)<14.2$) with stellar masses $log(M_*/M_\odot)>11.2$ are different. This is observed both in the local Universe, when comparing SDSS groups from \cite{2007ApJ...671..153Y} with field galaxies, and in our COSMOS group and field samples.  This is also true for massive central ellipticals, where models clearly predict a major impact from mergers  \cite{2011arXiv1105.6043S} . 

This result is at variance with predictions from the standard hierarchical model (e.g., from \cite{2011arXiv1105.6043S} ), which predicts instead that the mass-normalized radius in our group mass range should be $\approx$~2 times larger than in the field.  When taking into account uncertainties due our sample size, photometric redshift estimates and halo masses, this prediction remains, although the effect is reduced where our galaxy sample size is limited.  Model predictions are consistent with our COSMOS results at 1~$\sigma$, mainly because of the sample size. However, for our SDSS sample, the difference between the observed and predicted mass-normalized radii is at variance at $> 3 \sigma$. While the observations do not show any dependence of the mass-normalized radius with environment, the Shankar et al. (2012) predicts that galaxy mass-normalized radii over our group mass range should be $\sim 1.5$ larger than that in the field.
In future work, we will investigate the dependence of this result on the specific model (Shankar et al., in preparation) and on the properties of SDSS sample (Huertas--Comany et al., in preparation) used here. 


\item BGGs and satellite galaxies of similar stellar mass ($log(M_*/M_\odot)>11$) evolve in a similar way. This is in disagreement with hierarchical models, that also predict a difference between centrals and satellites living in halos with similar masses than our groups ($log(M_*/M_\odot)\sim13.4-13.6$). In fact, Shankar et al. (2011) has shown that galaxies residing at the centre of more massive haloes should, at fixed stellar mass, experience more mergers and thus be larger than their counterparts in less massive haloes.

\end{itemize}

\end{enumerate}

Future work will involve a detailed detailed discussion of environmental effects on models (Shankar et al. 2012, in prep) as well as an extension to higher redshifts by studying the sizes of ETGs in distant clusters of galaxies (Delaye et al. 2012, in prep).

\vspace{2cm}

\thanks{
The authors thank Boris Haussler for his great help with \textsc{galapagos} and M.R. George for putting together the main catalog used in this paper. We are also thankful to P. Capak, O. Ilbert as well as to K. Bundy and A. Leauthaud for interesting discussions. FS acknowledges support from a Marie Curie grant. 
}

\bibliographystyle{aa}
\bibliography{biblio_MNRAS}

\clearpage

\appendix

\section{Visual morphological classifications}
\label{app:morpho}
Some results presented in this paper are based on an automated morphological classification of galaxies at high redshift. Despite of the extensive tests performed to properly assess the accuracy of our classification, the separation of ellipticals from lenticulars might still remain a challenge.
In this appendix, we want to make sure that our results do not change while using an alternative classical visual classification. Two of us (SM and MHC) performed independent detailed visual classifications of all passive galaxies previously selected as early-type by galSVM (we assume that the separation between early and late is good enough). SM classified galaxies into three morphological classes (Ellipticals, Lenticulars and early-spirals) while MHC followed a slightly different criterion and separated galaxies into two classes (disk, no disk).
In Figs.~\ref{fig:morpho_compare} and \ref{fig:morpho_compare2} we reproduce figures~\ref{fig:evol_obser} and \ref{fig:evol_ETG} of the main text using the 3 morphological classifications (2 visual and 1 automated). While there are some differences between the three estimates, the main trends discussed in the text do hold and the results are consistent within the error bars. This confirms that our morphological classification is robust and that the major results presented in the paper are not biased by the automated classification. Most of the differences between visual and automated classifications are seen in the low mass elliptical population for which visual classifications show a steeper evolution than that estimated while using an automated classification.  When the evolution of low mass ellipticals is compared to their massive counterpart, it is still less steep as discussed in section~\ref{sec:ells} and shown in table~\ref{tbl:alpha_values}.

\begin{figure*}
\includegraphics[scale=0.75]{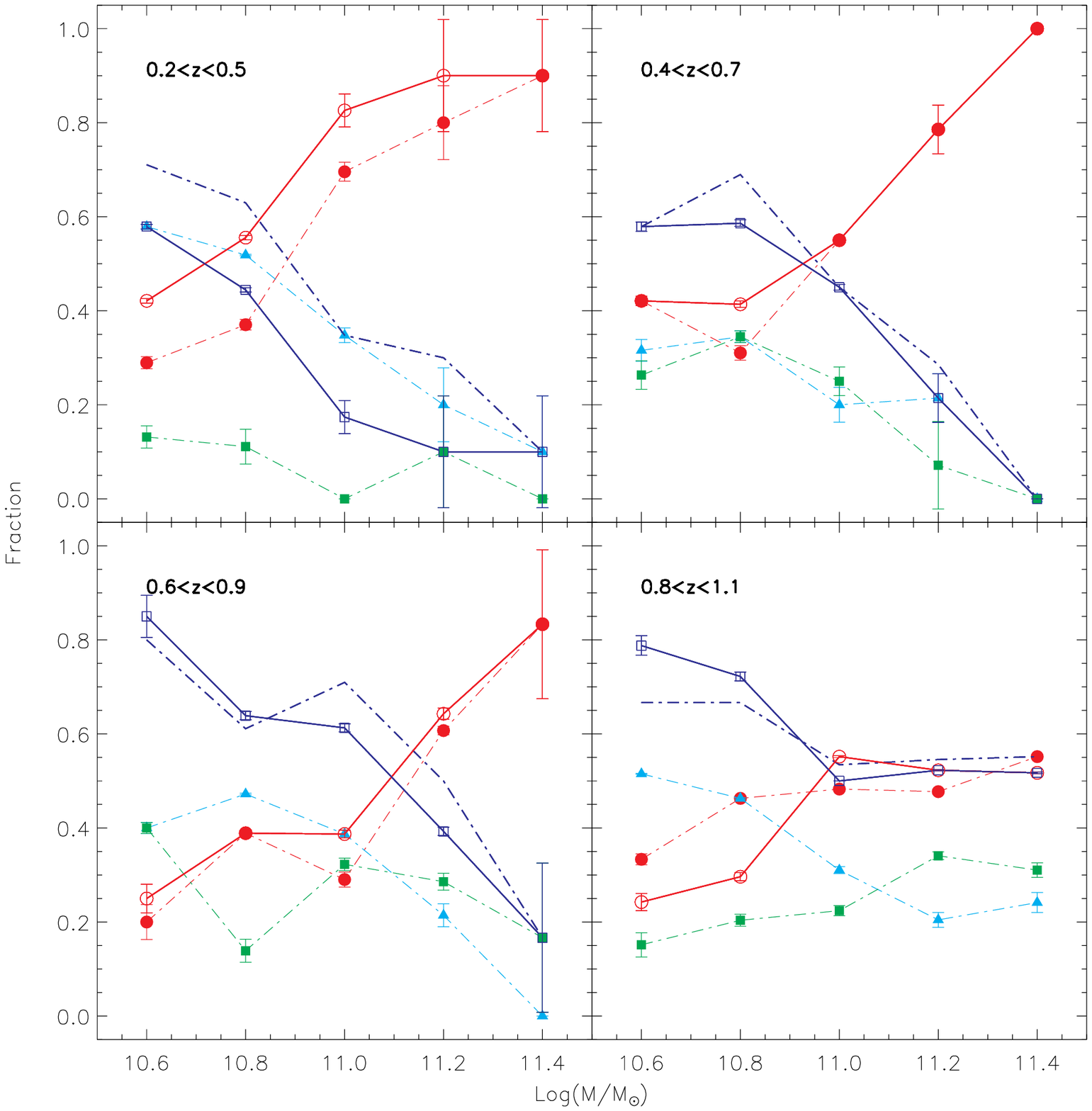}
\caption{Morphological fractions of passive galaxies as a function of stellar mass and redshift using the visual morphological classifications by MHC (solid lines) and SM (dotted-dashed lines). For MHC: red solid lines are ellipticals and blue are galaxies with an observed disk component. For SM: green lines are Sas, cyan lines are S0s and red lines are ellipticals. The dashed-dotted blue line shows the fraction of galaxies with an observed disk component (S0s+Sas). The uncertainties are calculated following Gehrels (1986; see Section 3 for binomial statistics; see also Mei et
al. 2009). These approximations apply even when ratios of
different events are calculated from small numbers, 
and yield the lower and upper limits of a binomial distribution 
within the 84\% confidence limit, corresponding to 1$\sigma$.  }
\label{fig:morpho_compare}
\end{figure*}


\begin{figure*}
\includegraphics[scale=0.75]{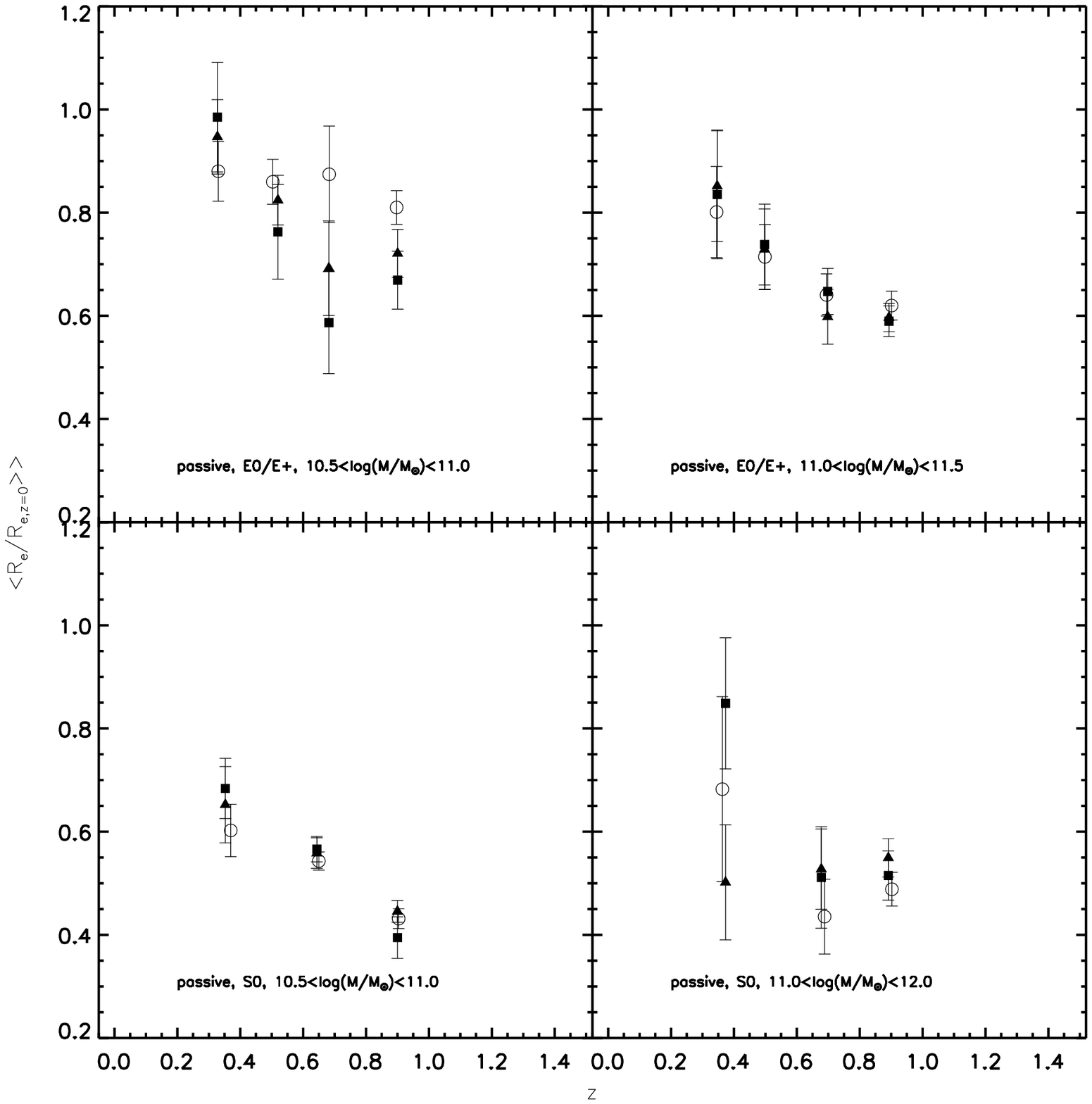}
\caption{Size evolution of passive ellipticals (top row) and S0s (bottom row) in two different mass bins using three different morphological classifications (see text for details). Empty circles show the automated classification from galSVM, squares are the visual classification from SM and triangles indicate are obtained with the visual classification of MHC. }
\label{fig:morpho_compare}
\end{figure*}

\begin{figure*}
\includegraphics[scale=0.7]{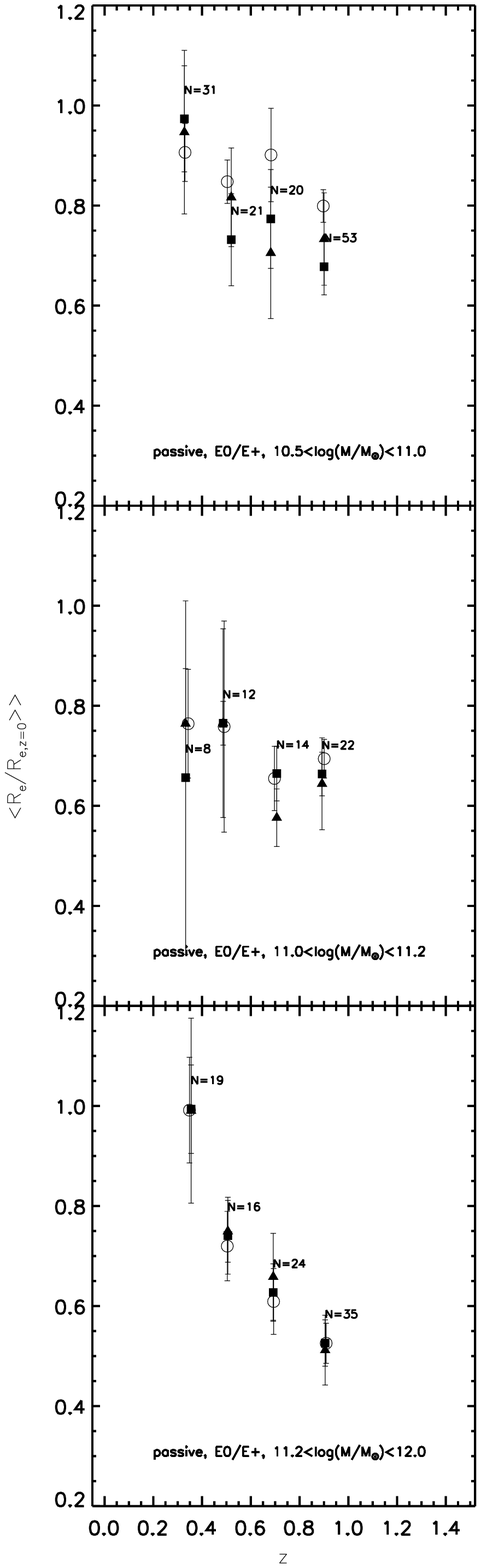}
\caption{Size evolution of elliptical (left column) passive galaxies in three mass bins using three different morphological classifications (see text for details). Empty circles show the automated classification from galSVM, squares are the visual classification from SM and triangles indicate are obtained with the visual classification of MHC. Top row: $10.5<log(M/M_\odot)<11$, middle row: $11.0<log(M/M_\odot)>11.2$ and bottom row: $11.2<log(M/M_\odot)>12$. Numbers indicate the number of objects in each redshift bin. Error bars are the scatter of the distributions. }
\label{fig:morpho_compare2}
\end{figure*}

\begin{table*}
\begin{tabular}{ccc}
\hline\hline\noalign{\smallskip}
Stellar mass bin & Morphology & $\alpha$\\ 
\hline\hline\noalign{\smallskip}
$10.5<log(M/M_\odot)<11$ & galSVM & $0.34\pm0.17$\\
 &   MHC & $0.52\pm0.23$\\
 &  SM & $0.59\pm0.27$\\
 $11<log(M/M_\odot)<11.2$ & galSVM & $0.63\pm0.21$\\
 &   MHC & $0.85\pm0.22$\\
 &  SM & $0.61\pm0.23$\\
 $11.2<log(M/M_\odot)<12$ & galSVM & $0.98\pm0.21$\\
 &   MHC & $0.94\pm0.21$\\
 &  SM & $0.95\pm0.21$\\
 
\end{tabular}
\caption{$\alpha$ values parameterizing the size evolution of ellipticals ($r_e\propto(1+z)^{-\alpha}$) in three mass bins obtained using three different morphological classifications.}.
\label{tbl:alpha_values}
\end{table*}

\section{Environment and galaxy selection}
\label{app:close}

We include here the equivalent of Fig.~\ref{fig:size_ratios} but only with galaxies having a probability $P_{MEM}>0.8$ to be a group member and being at a distance $d<0.5\times R_{200}$ of the cluster center. Our results do not change and remain basically the same as the ones shown in table~\ref{tbl:stats}. The size distribution in mass and redshift bins does not depend on environment.

\begin{figure*}
\includegraphics[scale=0.7]{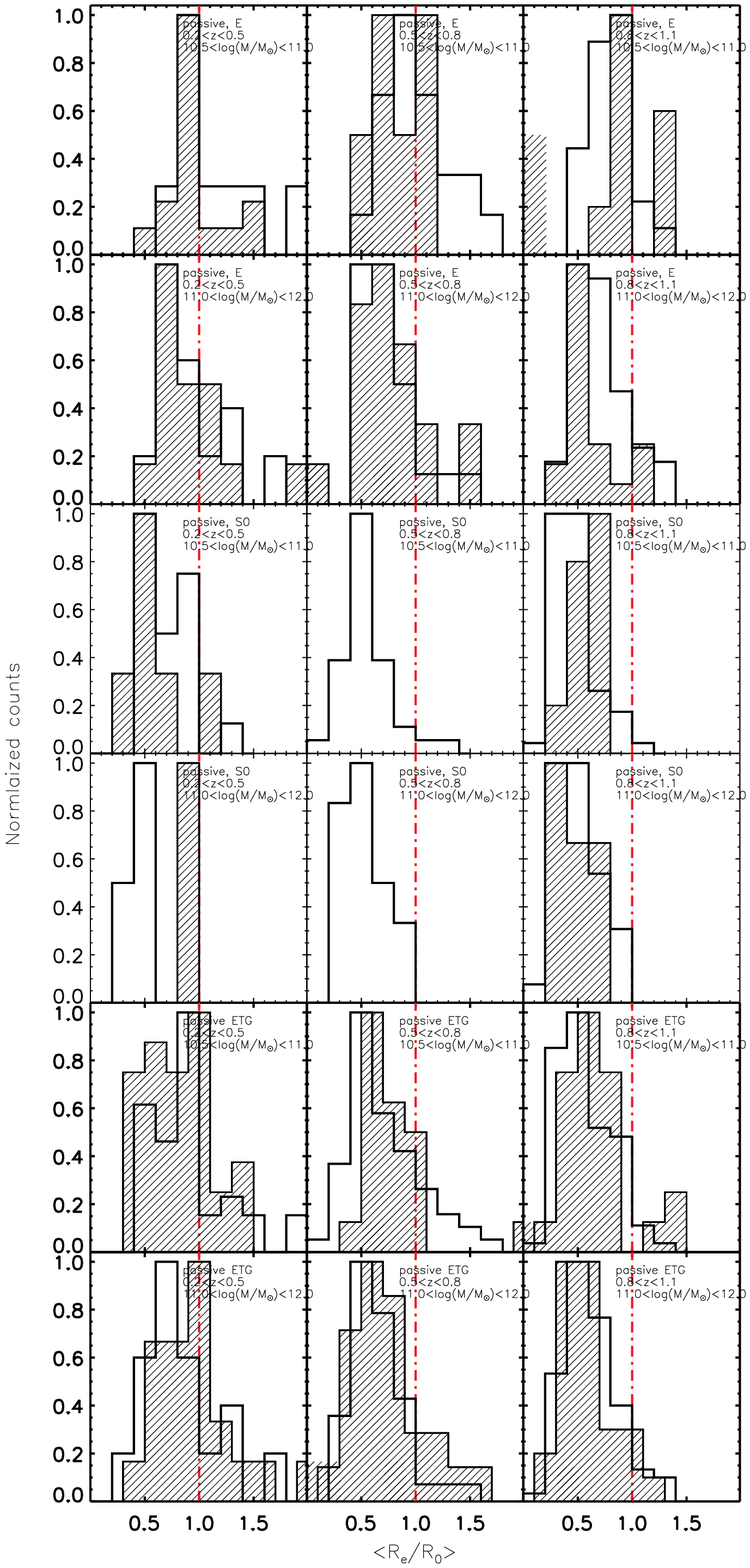}
\caption{Same as figure~\ref{fig:size_ratios} but only for galaxies having a probability $P_{MEM}>0.8$ to be a group member and being at a distance $d<0.5\times R_{200}$ of the cluster center. Wherever only one histogram is shown it is because there are too few objects in that bin.  }
\label{fig:size_ratios_close}
\end{figure*}

\clearpage





\label{lastpage}

\end{document}